\documentclass[11pt, a4paper]{article}
\usepackage{jcappub}

\usepackage{booktabs}
\usepackage{multirow}
\usepackage{amssymb}
\usepackage[export]{adjustbox}
\usepackage{floatrow}
\newfloatcommand{capbtabbox}{table}[][3.5in]
\newfloatcommand{capbfigbox}{figure}[][2.3in]
\usepackage{blindtext}
\usepackage{amsmath}
\usepackage{booktabs}
\usepackage{aas_macros}

\newcommand{\Fermi}{\textit{Fermi}}

\begin{document}

\hspace*{110mm}{\large \tt FERMILAB-PUB-15-255-A}

\vskip 0.2in

\title{The Galactic Center GeV Excess from a Series of Leptonic Cosmic-Ray Outbursts}

\author[a]{Ilias Cholis}
\emailAdd{cholis@fnal.gov}

\author[b]{Carmelo Evoli}
\emailAdd{carmelo.evoli@desy.de}

\author[c]{Francesca Calore}
\emailAdd{f.calore@uva.nl}

\author[d]{Tim Linden}
\emailAdd{trlinden@uchicago.edu}

\author[c]{Christoph Weniger}
\emailAdd{c.weniger@uva.nl}

\author[a,e]{and Dan Hooper}
\emailAdd{dhooper@fnal.gov}

\affiliation[a]{Fermi National Accelerator Laboratory, Center for Particle
Astrophysics, Batavia, IL 60510, USA}
\affiliation[b]{Instit{\"u}t f{\"u}r Theoretische Physik, Universit{\"a}t Hamburg, Luruper Chaussee 149, D-22761, Hamburg, Germany}
\affiliation[c]{GRAPPA, University of Amsterdam, Science Park 904, 1090 GL
Amsterdam, Netherlands}
\affiliation[d]{University of Chicago, Kavli Institute for Cosmological Physics, 933 E. 56th St., Chicago, IL 60637 USA}
\affiliation[e]{University of Chicago, Department of Astronomy and Astrophysics, 5640 S. Ellis Ave., Chicago, IL 60637 USA}

\abstract{
It has been proposed that a recent outburst of cosmic-ray electrons could account for the excess of GeV-scale gamma rays observed from the region surrounding the Galactic Center. After studying this possibility in some detail, we identify scenarios in which a series of leptonic cosmic-ray outbursts could plausibly generate the observed excess. The morphology of the emission observed outside of $\sim1^{\circ}-2^{\circ}$ from the Galactic Center can be accommodated with two outbursts, one which took place approximately $\sim10^6$ years ago, and another (injecting only about 10\% as much energy as the first) about $\sim10^5$ years ago. The emission observed from the innermost $\sim1^{\circ}-2^{\circ}$ requires one or more additional recent outbursts and/or a contribution from a centrally concentrated population of unresolved millisecond pulsars. In order to produce a spectrum that is compatible with the measured excess (whose shape is approximately uniform over the region of the excess), the electrons from the older outburst must be injected with significantly greater average energy than those injected more recently, enabling their spectra to be similar after $\sim10^6$ years of energy losses. 
}

\maketitle

\section{Introduction}
\label{sec:intro} 

Over the past five years, an excess of GeV-scale gamma rays collected by the 
Large Area Telescope (LAT) aboard the \Fermi~satellite has been
reported from the direction of the region surrounding the Galactic
Center (GC)~\cite{Goodenough:2009gk, Hooper:2010mq, Hooper:2011ti, Abazajian:2012pn,
Gordon:2013vta, Hooper:2013rwa, Abazajian:2014fta, Daylan:2014rsa,
Calore:2014xka, fermigc}.\footnote{A similar residual was also mentioned by the \Fermi\ team in
2009~\cite{Vitale:2009hr}. No details were provided, however, regarding its spatial
morphology, systematic issues regarding astrophysical backgrounds, or
possible interpretations.}
The consistency of the spectrum, angular distribution, and overall normalization of the GeV excess with predictions of annihilating
dark matter (DM) has generated a great deal of interest. Specifically, the
signal has been shown to be distributed with approximate spherical symmetry
about the GC, with a profile that corresponds to a DM
density that scales as $\rho_{\rm DM} \propto r^{-1.2}$, compatible with expectations from simulations~\cite{Navarro:1995iw,Navarro:1996gj,Navarro:2008kc,Diemand:2008in,DiCintio:2014xia, DiCintio:2013qxa}. The
spectral shape of the excess is also in good agreement with that predicted from
annihilating DM (see, for example, Figure 2 of Ref.~\cite{Calore:2014nla}).  For the case of DM particles that annihilate 
predominately to $\bar{b}b$, the observed spectrum is consistent with a dark matter mass
in the range of 43--55 GeV at 68.3\% CL~\cite{Calore:2014nla} (or 36--51
GeV~\cite{Daylan:2014rsa}, 35--43 GeV~\cite{Abazajian:2014fta}, as found by the
authors of other recent analyses; see also Ref.~\cite{Agrawal:2014oha}).
Annihilations to lighter quarks (or to a combination of quark species) can also
provide a good fit, although for lower values of the DM mass~\cite{Calore:2014nla, Daylan:2014rsa}, whereas annihilations into $hh$, $W^+W^-$, $ZZ$, or $hZ$ final states would point to higher DM masses~\cite{Agrawal:2014oha, Calore:2014nla, Caron:2015wda}.

In addition to annihilating DM, two classes of astrophysical explanations have been
proposed for the GC GeV excess. The first posits that the signal
is generated by a large population of unresolved millisecond pulsars. 
This
possibility is motivated in large part by the spectral
similarity between such sources and the observed GC excess~\cite{Hooper:2010mq, Abazajian:2010zy}.
As recent analyses have shown the excess to be spatially extended to a radial
projected distance of
at least $\sim$$10^{\circ}$~\cite{Hooper:2013rwa, Daylan:2014rsa,
Calore:2014xka}, such an interpretation is constrained by the results of point source searches in the region. In particular, if enough millisecond pulsars were
present within the inner 1.8 kiloparsecs (kpc) of the Galaxy to account for the observed
emission, it was shown in Ref.~\cite{Cholis:2014lta} that there should exist $\sim$\,60 such sources with $L_{\gamma}>10^{35}$ erg/s (integrated above 0.1 GeV), corresponding to a flux that is significantly greater than many of the gamma-ray sources detected from the direction of the Inner Galaxy (see also Refs.~\cite{Hooper:2013nhl,Cholis:2014noa}). These and other works suggest that only up to 5--10\% of the GeV excess emission
could arise from an unresolved population of millisecond pulsars~\cite{Cholis:2014lta,Calore:2014oga,Hooper:2013nhl,
Cholis:2014noa}.  That being said, a debate is ongoing regarding how many
near-threshold gamma-ray point sources could be present in the Inner
Galaxy~\cite{BartelsInPrep,lisanti}. Alternatively, this constraint could potentially be evaded if the luminosity function of the millisecond pulsars in the region
surrounding the GC is significantly different from that observed
elsewhere in the Galaxy~\cite{Yuan:2014rca, Petrovic:2014xra,Cholis:2014lta}. Although young pulsars could also contribute, as suggested in Ref.~\cite{O'Leary:2015gfa}, the lack of excess emission from along the Galactic Plane strongly limits their role.

The second explanation proposes that the excess emission may be generated as a result of one or more
recent cosmic-ray outbursts. In principle, this could be the result of either
cosmic-ray protons interacting with gas via neutral pion production~\cite{Carlson:2014cwa}, or cosmic-ray electrons undergoing inverse
Compton scattering (ICS)~\cite{Petrovic:2014uda}. As hadronic scenarios predict a
gamma-ray signal that is significantly extended along the Galactic Plane and otherwise
correlated with the distribution of gas~\cite{Carlson:2014cwa}, they are highly
incompatible with the observed characteristics of the excess. A leptonic
outburst, in contrast, could plausibly lead to a signal that is more smoothly
distributed and spherically symmetric. In light of these considerations, we
consider a series of leptonic outbursts taking
place over the past $\sim$\,$10^6$ years to be the most plausible of the astrophysical
explanations proposed for the GC GeV excess.  Previous work, however, has found it to be difficult to simultaneously explain the spectrum and spatial morphology of the GC excess~\cite{Petrovic:2014uda}. In this paper, we revisit this possibility.
In particular, given the apparent self-similarity of the spectral shapes of
the excess emission across different regions of the Inner Galaxy~\cite{Calore:2014xka}, it is important to
re-evaluate whether and under what circumstances outburst scenarios could indeed
account for such an observation.

Our work extends the analysis of Ref.~\cite{Petrovic:2014uda} in a number of significant
ways. Firstly, we go beyond semi-analytical solutions to the diffusion
equation, allowing us to take into account spatially dependent energy
losses and variations in the cosmic-ray propagation conditions.  Secondly, we
consider the effects of re-acceleration, convection, and anisotropic diffusion, which were neglected previously.
Thirdly, we systematically perform fits to the entire Inner Galaxy, spanning the $40^\circ\times40^\circ$ region around the GC.  And finally, we perform a systematic scan over source and propagation parameters.

The present paper is organized as follows. In Sec.~\ref{sec:outbursts}, we present a 
framework for modeling outburst events in the GC and discuss the possible connection between such an event and the gamma-ray features known as the \textit{Fermi} Bubbles.
In Sec.~\ref{sec:model}, we describe the data sets used in 
our analyses and we provide technical details about the simulation of outburst models using
cosmic-ray propagation codes. We present our results in
Sec.~\ref{sec:results}, and summarize our conclusions in Sec.~\ref{sec:conclusions}.

\section{Cosmic-ray outbursts in the Inner Milky Way}
\label{sec:outbursts}

The modeling of the Galactic gamma-ray foregrounds and backgrounds generated by diffuse emission
processes in the Milky Way ($\pi^0$-production, inverse Compton scattering, and Bremsstrahlung),
is generally done under the assumption that the distribution of cosmic rays can be approximated by a steady-state solution. Diffuse emission models derived under this assumption have been 
unable to account for the observed characteristics of the GeV excess~\cite{Calore:2014xka, fermigc,Daylan:2014rsa}.

In some respects, the steady-state assumption appears to be well motivated. At present, the Milky Way's
supermassive black hole is in a quiescent
state~\cite{Genzel:2010zy}, and the star formation rate of the GC
is $\sim 0.1 \,M_{\odot}$/yr, near the average rate required to generate the
stellar population of the nuclear bulge over a period of $\sim 10^9$
years~\cite{Crocker:2011en}. There are a number of indications, however, that
the intensity of cosmic-ray emission from the GC region may be far from uniform
in time, instead consisting in part of a series of energetic
outbursts~\cite{Paumard:2006im, Law:2009wv, Sofue:2003xm, Ponti:2012pn,
Su:2010qj}, related either to outflows from the central black hole or to nuclear starburst events. 

The strongest indication in favor of such a scenario is represented by the gamma-ray features known as the \textit{Fermi}
Bubbles~\cite{Dobler:2009xz,Su:2010qj,Fermi-LAT:2014sfa}.  The diffuse
gamma-ray emission associated with the Bubbles exhibits an hourglass-like morphology, extending up to
$\sim$ $50^{\circ}$ ($\sim$10 kpc) above and below the Galactic Plane, and corresponds to a total gamma-ray luminosity of $(0.3-1) \times 10^{38}$ erg/s \cite{Malyshev:2010xc, :2012rta}.  
The spectrum of the gamma-ray emission from the Bubbles is quite hard, $dN/dE_{\gamma} \propto E_{\gamma}^{-2}$
 between $E_{\gamma} \sim $\,1 -- 100 GeV, and extends up to at least several hundred GeV. The spectrum does not show any significant variation across the Bubbles region.

 Both leptonic and hadronic models have been proposed to explain the \textit{Fermi} Bubbles. Leptonic models are particularly attractive due to the fact that the same population of cosmic-ray electrons could generate the gamma-ray Bubbles through ICS, as well as the microwave ``haze'' observed by \textit{WMAP} and \textit{Planck}~\cite{Finkbeiner:2003im, Dobler:2007eg, Dobler:2011rd, :2012rta} via synchrotron emission. 
Intensity, spatial and spectral correlations between the gamma-ray and microwave signals have been shown to support this 
hypothesis~\cite{Dobler:2009xz, :2012rta,  Dobler:2011rd, Hooper:2013rwa}. Various authors have proposed mechanisms for the origin of these energetic electrons, including, for example, recent ($\sim 1$--$3$ Myr ago) AGN jet activity 
from the central massive black hole~\cite{Guo:2011eg, Guo:2011ip, Yang:2012fy}, 
a spherical outflow~\cite{Zubovas:2011py}, a sequence of
shocks resulting from several accretion
events (involving first order Fermi acceleration)~\cite{Cheng:2011xd}, the stochastic acceleration of high
energy electrons by large-scale turbulence taking place throughout the volume
of the Bubbles (involving second order Fermi acceleration) \cite{Mertsch:2011es}, and even the annihilations of TeV-scale DM 
into leptons~\cite{Cholis:2009va, Dobler:2011mk}.

Alternatively, hadronic models for the \textit{Fermi} Bubbles have been
proposed~\cite{Crocker:2010dg}, associated with long time scale ($\sim$Gyr) star formation 
in the GC, transferred away from the Galactic Plane by strong winds. Notably, this class of models predicts a large 
neutrino counterpart signal that could be probed by current and future neutrino telescopes \cite{Cholis:2012fr, Lunardini:2011br}, and that may already be in some tension with high energy ($\mathcal{O}$(30 -- 1000) TeV) neutrino flux measurements~\cite{Ahlers:2015moa}.


Motivated by the possible connection with the \textit{Fermi} Bubbles, we focus in this work on the case of cosmic-ray electron outbursts, from the inner parsecs or tens of parsecs around the GC.  A challenge in connecting the GC GeV excess with the \textit{Fermi} Bubbles is that the GeV excess is approximately spherically symmetric with respect to the GC, whereas the Bubbles are strongly oriented perpendicular to the Galactic Plane. A common origin for these signals would require a scenario in which the $\sim 10$ GeV electrons responsible for the GeV excess diffuse highly isotropically, whereas the higher energy ($\sim10^2-10^3$ GeV) electrons responsible for the Bubbles propagate more rapidly and preferentially away from the Galactic Plane. Although we do not have a concrete proposal to account for these behaviors, one could imagine a scenario in which the magnetic fields of the GC efficiently isotropize the diffusion of low-energy electrons, while allowing higher energy particles to propagate and escape the region more freely. Alternatively, an outburst responsible for the \textit{Fermi} Bubbles may have been forced by the surrounding environment to propagate away from the Galactic Plane, but left the GC region in a state that allowed the electrons of subsequent outbursts to propagate more isotropically. For example, a large outburst of electrons injected with a jet-like morphology a few Myrs ago, may have not experienced strong diffusive re-acceleration within the inner tens of parsecs, allowing the electrons to maintain their original orientation and propagate preferentially away from the Galactic Plane while suffering relatively little energy losses. On the other hand, later outbursts, taking place $\sim$\,$0.1-1$ Myr ago, may have occurred under rather different conditions, including greater turbulence and stronger re-acceleration as the result of the earlier outburst. These more recent outbursts could thus propagate more isotropically, and experience more rapid synchrotron energy losses, leading to a morphology and spectrum that is similar to that of the GC GeV excess.

Cosmic-ray outbursts may have occurred at different times in the recent history of the Milky Way, each with different luminosities, persisting over different timescales, and injecting cosmic rays into the interstellar medium with different spectra. This allows us a great deal of freedom ({\it i.e.} many free parameters) to fit the morphology and spectrum of the GC GeV excess. In the following section, we describe our modeling of cosmic-ray outbursts in an attempt to generate the observed characteristics of the GC GeV excess.


\section{Modeling and data analysis}
\label{sec:model}

\subsection{Gamma-ray data used in this work}

In searches for DM annihilation products from the Inner Galaxy, the regions of optimal sensitivity (maximal signal-to-background) can extend up to tens of degrees away from the disk, with the details
depending on the specific emission profile~\cite{Serpico:2008ga, Cholis:2009gv, Bringmann:2011ye,Weniger:2012tx,
Nezri:2012xu, Tavakoli:2013zva}. Several analyses of the GC GeV excess have focused on what is known as the ``Inner Galaxy" -- a region including the surrounding tens of degrees around the GC and excluding portions of the Galactic Plane (where the backgrounds are highest). The fact that the GeV excess is observed to extend to relatively high latitudes has been interpreted in support of a DM interpretation of this signal~\cite{Hooper:2013rwa, Huang:2013pda, Daylan:2014rsa,Calore:2014xka}, as has its approximate spherical symmetry with respect to the GC~\cite{Daylan:2014rsa,Calore:2014xka}. In contrast, many proposed astrophysical explanations for the GeV excess predict emission that is localized preferentially around the disk~\cite{Carlson:2014cwa,O'Leary:2015gfa}.

In this study, we make primary use of the results of Ref.~\cite{Calore:2014xka}, which characterized the spectrum and morphology of the GeV excess over the region of the Inner Galaxy. When discussing the region of the GC and Galactic Plane (which are excluded from the Inner Galaxy studied in Ref.~\cite{Calore:2014xka}), we draw from the results of Ref.~\cite{Daylan:2014rsa}. In the following subsections, we describe separately the adopted data sets. 

\subsubsection{The Inner Galaxy} 
For the Inner Galaxy analysis, we adopt the GeV excess data as derived in Ref.~\cite{Calore:2014xka}.
The excess was characterized within the region $|\ell|<20^\circ$ and $2^\circ<|b|<20^\circ$, thus avoiding the critical 
few degrees in latitude where point-source subtraction and modeling of the diffuse Galactic gamma-ray 
emission are most difficult. 
Ref.~\cite{Calore:2014xka} derived the spectral properties of the excess in the
main region-of-interest by removing Galactic foregrounds and backgrounds.  Care was
exercised in estimating the typical uncertainties of this removal by analyzing
residuals along the Galactic Disk.
The corresponding systematic uncertainties are represented by a full
covariance matrix that encodes how uncertainties in the normalization and
spectral slope of the Galactic diffuse foreground components affect the determination of the GeV excess
spectrum.
In addition to determining the spectral properties of the excess in the main region-of-interest, 
Ref.~\cite{Calore:2014xka} also characterized the morphology of the signal in several sub-regions,
probing the emission spectrum up to $|b| = 10^{\circ}$--$20^\circ$, as well as its morphological properties. 
Exploiting the full spectral and morphological properties of the GeV excess emission
is crucial for shedding light onto the origin of the excess emission, and it leads to tighter constraints than would result from considering only 
the spectrum from within the main region-of-interest.

In our analysis, we perform fits to the spectrum of the GC GeV
excess as it was extracted in Ref.~\cite{Calore:2014xka}. In particular, we will use the fluxes from the ten segmented regions
within $|\ell|<20^\circ$ and $2^\circ<|b|<20^\circ$, from energies of 300 MeV
to 500 GeV.  When calculating the quality of the fits, we take into account the
full covariance of the systematic errors~\cite{Calore:2014xka}.  Namely, the
$\chi^2$ function is given by:
\begin{equation}
    \chi^2 = \sum_{i=1}^{10} \sum_{j,k=1}^{24} 
    (d_{ij} - \mu_{ij} ) (\Sigma_{jk}^{i})^{-1}
    (d_{ik} - \mu_{ik} )\;,
\end{equation}
where $d_{ij}$ ($\mu_{ij}$) denotes the measured (predicted) flux in region $i$
and energy bin $j$, and $\Sigma_{jk}^i$ is the covariance matrix for energy
bins $j$ and $k$ in region $i$.  Note that, as in Ref.~\cite{Calore:2014xka},
we neglect the covariance of the excess emission between different regions.
When quoting $p$-values below, we always refer to the value of this $\chi^2$
function for the ten region fit, and assume it follows a $\chi^2_k$
distribution with $k=240-1$ degrees of freedom.

For a given source and cosmic-ray diffusion setup, we simulate the excess
emission expected in the ten different regions described below, up to an
overall normalization which depends on the injected energy.  This normalization
is then obtained by a fit to the data, minimizing the above $\chi^2$
function.

\subsubsection{The Galactic Center}
 
Explanations of the GeV excess must also confront the bright gamma-ray emission observed from the region within $1^{\circ}-2^\circ$ of the GC, which has been shown to be radially extended, spherically symmetric, centered within $0.05^{\circ}$ of the GC, and have a spectrum consistent with the full extent of the gamma-ray excess~\citep{Daylan:2014rsa}. When considering the emission from the region immediately surrounding the GC (subsection~\ref{gcsection}), or from along the Galactic Plane (subsection~\ref{gpsection}), we will use the results of the analysis presented in Ref.~\citep{Daylan:2014rsa}.

\subsection{Numerical simulations of outburst events}

To simulate the propagation of electrons in the GC and Inner Galaxy regions, we make use of the publicly available numerical codes \texttt{Galprop}~\cite{Galprop} and \texttt{DRAGON}~\cite{DiBernardo:2009ku,Evoli:2008dv}. Numerical codes are necessary to properly model energy losses in this region, since they depend on the distributions of gas, interstellar radiation field, and magnetic field.

\subsubsection{Gamma-ray diffuse emission from outbursts using \texttt{Galprop}}
\label{sec:Burst_Galprop}

The GC GeV excess is discernible from throughout the innermost $\sim$2 kpc of the Milky Way. In this part of the Galaxy, cosmic-ray propagation conditions can be very different from those found locally, which are constrained by measurements of cosmic-ray secondary-to-primary ratios ({\it e.g.} boron-to-carbon). Turning this around, the propagation conditions in the Inner Galaxy are only weakly constrained by local cosmic-ray measurements, allowing us considerable freedom in selecting the parameters that describe diffusion, re-acceleration, convection, and energy loss processes for cosmic rays in the Inner Galaxy. We calculate the spectrum of gamma rays produced through either ICS with low-energy photons~\cite{Jones:1968zza, Blumenthal:1970gc}, or through interactions with interstellar gas, giving rise to Bremsstrahlung emission~\cite{Koch:1959zz, 1969PhRv..185...72G, Blumenthal:1970gc}.

In our simulations, we assume that the cosmic-ray electron outbursts occur homogeneously within a cylinder of
50 pc radius and 100 pc height, centered at the GC, and with a spectrum that takes the form of a power-law with an exponential cutoff:
\begin{equation}
  \frac{dN_{e}}{dE_{e}} = \mathcal{N}\, E_{e}^{-\alpha} \, \exp\{ -E_{e}/E_{\rm cut}\} .
    \label{eqn:Injection}
\end{equation}
We allow the $\mathcal{N}$ to vary freely, keeping in mind that the current power in gamma-ray emission from the \textit {Fermi} Bubbles multiplied by an estimated age of $10^6$ years suggests a total energy of order $\sim 10^{51}$ erg.  We allow the spectral index to take on values in the range of $\alpha =1-3$ and consider cutoff energies of 15 to 100 GeV (for single burst models simulated with \texttt{DRAGON}, we also consider $E_{\rm cut}=1$ TeV). 
We clarify that this injection spectrum denotes the spectrum of electrons at $\sim50$ pc from the GC, and not necessarily
as they were injected from their original sources (collections of supernovae or the central black hole).
This allows us to consider values such as $\alpha \sim 1$ which are not well motivated by Fermi acceleration.
As discussed in section~\ref{sec:outbursts}, strong re-acceleration and 
rapid synchrotron cooling within the inner tens of pc of the Galaxy could lead to a hardening of the spectral power-law of cosmic ray electrons 
combined with a spectral cutoff.  We note that previous work on leptonic outbursts assumed a power-law spectral shape~\cite{Petrovic:2014uda}, without an exponential cutoff.

To model the propagation of cosmic-ray electrons, we use the standard \texttt{Galprop} setup. The timescale that cosmic rays of a given rigidity remain within the region of the Inner Galaxy is set by the processes of diffusion and convection. For isotropic diffusion, the diffusion coefficient is given as a function of rigidity, $R$, by:
\begin{equation}
    D(R) \equiv D_{xx}(R) = D_{0} \left(\frac{R}{3 \, GV}\right)^{\delta}\,,
    \label{eqn:Diffusion}
\end{equation}
where $D_0$ is the diffusion coefficient at 3 GV and $\delta$ is the diffusion
index. We assume that any convection is perpendicular to the Galactic Disk, with a constant gradient $dv_{c}/dz$. 

Cosmic ray electrons suffer from energy losses due to synchrotron, ICS, and Bremsstrah-lung emission. For the Galactic magnetic field (which determines the rate of synchrotron energy losses), we assume a cylindrical symmetry with the following parameterization:
\begin{equation}
    B(r,z) = B_{0}\, e^{-r/r_c} \, e^{-|z|/z_{c}} \,,
    \label{eqn:B-field}
\end{equation}
where $B_0$ is the magnetic field at the center of the Galaxy and $r_c$ and $z_c$ are the characteristic scale lengths. For the interstellar radiation field (which determines the rate of ICS), we use the model employed within \texttt{Galprop v54}  (see Ref.~\cite{GALPROPSite}) but allow for the normalization of the emission from stars and dust grains to freely vary within a factor of 3 from the reference assumptions, while the contribution from the cosmic microwave background is kept fixed. For Bremsstrahlung off interstellar atomic hydrogen (HI) and molecular hydrogen ($H_{2}$) gas, we adopt the gas distributions used within \texttt{Galprop v54}~\cite{GALPROPSite}.

Cosmic ray electrons with energies below a few GeV can also also be diffusively re-accelerated. Diffusive re-acceleration is connected to spatial diffusion through the following relationship:
\begin{equation}
    D_{pp}(R) =  \frac{4}{3 \delta (2-\delta)(4-\delta)(2+\delta)} \frac{R^{2} v_{A}^{2}}{D_{xx}(R)} \, ,
    \label{eqn:Reacceleration}
\end{equation}
where $v_{A}$ is the Alfv$\acute{\textrm{e}}$n speed \cite{GALPROPSite, 1994ApJ...431..705S}.

In order to generate maps of the gamma-ray emission from an outburst of a given age, $\tau$, and duration, $\Delta\tau$, we  
first inject cosmic-ray electrons from the GC and propagate them until $\tau + \Delta\tau/2$ and then again inject the same spectrum of cosmic-ray electrons and propagate them until $\tau - \Delta\tau/2$, and then finally take the difference of these two maps. We have examined the convergence of these results for both the assumed spatial grid and the timestep employed in this work.
We take $\Delta\tau$ $= 1 \times 10^{4}$ yr for all outbursts, have assumed a cylindrical grid of $\Delta r = 50$ pc, $\Delta z = 50$ pc, and have set the \texttt{Galprop} timestep parameters to \texttt{start\_timestep} $= 5 \times 10^{3}$ yr,  \texttt{end\_timestep} $= 5$ yr and 
\texttt{timestep\_factor} $= 0.25$. 

In our analysis, we make use of the \texttt{Galprop v54} code~\cite{Strong:2007nh, GALPROPSite, GalpropV54} to produce ICS and Bremsstrahlung gamma-ray templates for the Inner Galaxy at various energies. We study a wide variety of outburst injection and propagation assumptions in the Inner Galaxy (for details, see Appendix~\ref{app:35models}).

\subsubsection{Modeling with \texttt{DRAGON}}
\label{sec:Burst_Dragon}

In addition to \texttt{Galprop}, we make use of the \texttt{DRAGON} code in our study to test outburst scenarios in which propagation proceeds inhomogeneously and/or anisotropically (which we parameterize using two diffusion coefficients, $D_{xx}$ and $D_{zz}$, which account for diffusion in the directions parallel to and perpendicular to the Galactic Plane, respectively.). Furthermore, we find that \texttt{DRAGON} performs better in terms of memory management and required computation time than the publicly available version of \texttt{Galprop}, 
allowing us to run the multi-dimensional parameter scan, as discussed in
the next section, in a reasonable amount of time.
Finally, the use of \texttt{DRAGON} allows us to test different assumptions for the gas distributions in the Inner Galaxy.

To simulate an outburst using \texttt{DRAGON}, we assume as an initial condition the analytic solution derived in Ref.~\cite{pulsars} for spatially constant energy losses evaluated at the time $t = 10^4$~years (which is much shorter than the outburst's age). We then solve the time-dependent diffusion equation for a total time corresponding to the age of the outburst, $\tau$. We adopt a spatial grid step size of 50~pc and a constant time step of 1~kyr. 


While the GeV excess appears to be spherically symmetric with respect to the GC, this symmetry is broken in outburst scenarios to some degree by energy losses that are highest near or within the Galactic Plane. This is primarily the result of Bremsstrahlung, whereas ICS occurs comparatively uniformly in space. By introducing anisotropic diffusion ($D_{zz} \ne D_{xx}$), we hope to identify scenarios that predict a more spherically symmetric gamma-ray signal from a given cosmic-ray outburst.

\subsection{Parameter inference for a single outburst}
\label{sec:Multinest}

\begin{table}
\begin{tabular}{lccc}
\toprule
Parameter & Units & Range & Prior \\
\midrule
$\alpha$ & & 1--3       & lin \\
$\delta$ & & 0.1--1.0 & lin \\
$D_0$   & $\rm 10^{28}\,cm^2/s$ & 0.1--20 & lin \\
$D_{zz}/D_{xx}$ & & 0.1--10 & log \\
$v_A$   &$\rm km/s$ & 0--200 & lin\\
$\tau$ & $\rm Myr$ & 0.1--5 & lin \\
\bottomrule
\end{tabular}
\caption{The range of single outburst model parameters considered in our scans. For each parameter, we indicate whether we adopt logarithmic or linear priors in the Bayesian parameter scan.}
\label{tab:ranges}
\end{table}

In order to efficiently scan through the multi-dimensional space of propagation
and source parameters (except for the overall normalization, which we obtain in the
final $\chi^2$ minimization), we make use of the nested sampling algorithm
implemented in \texttt{MultiNest}~\cite{Feroz:2007kg, Feroz:2008xx, Feroz:2013hea}.  We
have directly coupled \texttt{MultiNest} to the cosmic-ray propagation code \texttt{DRAGON}.

Convergence is usually obtained after $\mathcal{O}(10^4)$ evaluations.  Since
some of the propagation and source parameters are either not well constrained
or are highly degenerate, the choice of priors is critical for parameter
estimation (although it does not affect our best-fit scenarios, which are always
robustly identified).  Our choice of priors (either logarithmic or linear) for the
different parameters in the scans is summarized in Tab.~\ref{tab:ranges}. Note
that in this section, we fix the energy cutoff to a high value of $E_{\rm cut} =1\ \rm TeV$, but will later consider lower values for this quantity within the context of two outburst models.

\texttt{MultiNest} provides as output a list of parameter points, with the density of
points being a proxy for the posterior probability distribution function.
The latter is given by Bayes' theorem:
\begin{equation}
    {\rm post}(\vec\theta) \propto \mathcal{L}(\vec\theta| \vec d ) {\rm prior}(\vec\theta),
\end{equation}
where $\vec\theta$ and $\vec d$ denote all model parameters and
data, respectively, $\rm post(\vec\theta)$ and $\rm prior(\vec\theta)$ denote
the posterior and prior probability distribution functions, and the profiled likelihood function is given by:
\begin{equation}
    -2\ln\mathcal{L}(\vec\theta)= \min_{\rm norm} \chi^2\;.
\end{equation}
As mentioned above, we allow the overall normalization of the injected excess to vary freely in each parameter evaluation.
When making statements about excess
energetics, we always take into account the uncertainty in the fitted normalization.

\section{Results}
\label{sec:results}

In this section, we first present our results from the \texttt{MultiNest} approach using \texttt{DRAGON}, 
(see Secs.~\ref{sec:Burst_Dragon} and~\ref{sec:Multinest}). We will show in
subsection~\ref{Sec:single}
that models with a single outburst are unable to explain the overall morphology of the GC excess.
In subsection~\ref{sec:2bursts} we consider models with two outbursts of different ages, which enable us to obtain reasonable fits to the gamma-ray data over most of the Inner Galaxy (neglecting for the moment the region within $2^{\circ}$ of the Galactic Plane). Finally, in subsection~\ref{sec:morphology}, we discuss the morphology of the gamma-ray emission from the regions within the innermost $2^{\circ}$ around the GC and along the Galactic Plane.

\subsection{Single outburst model}\label{Sec:single}

\begin{figure}
\begin{center}
\includegraphics[width=0.47\linewidth]{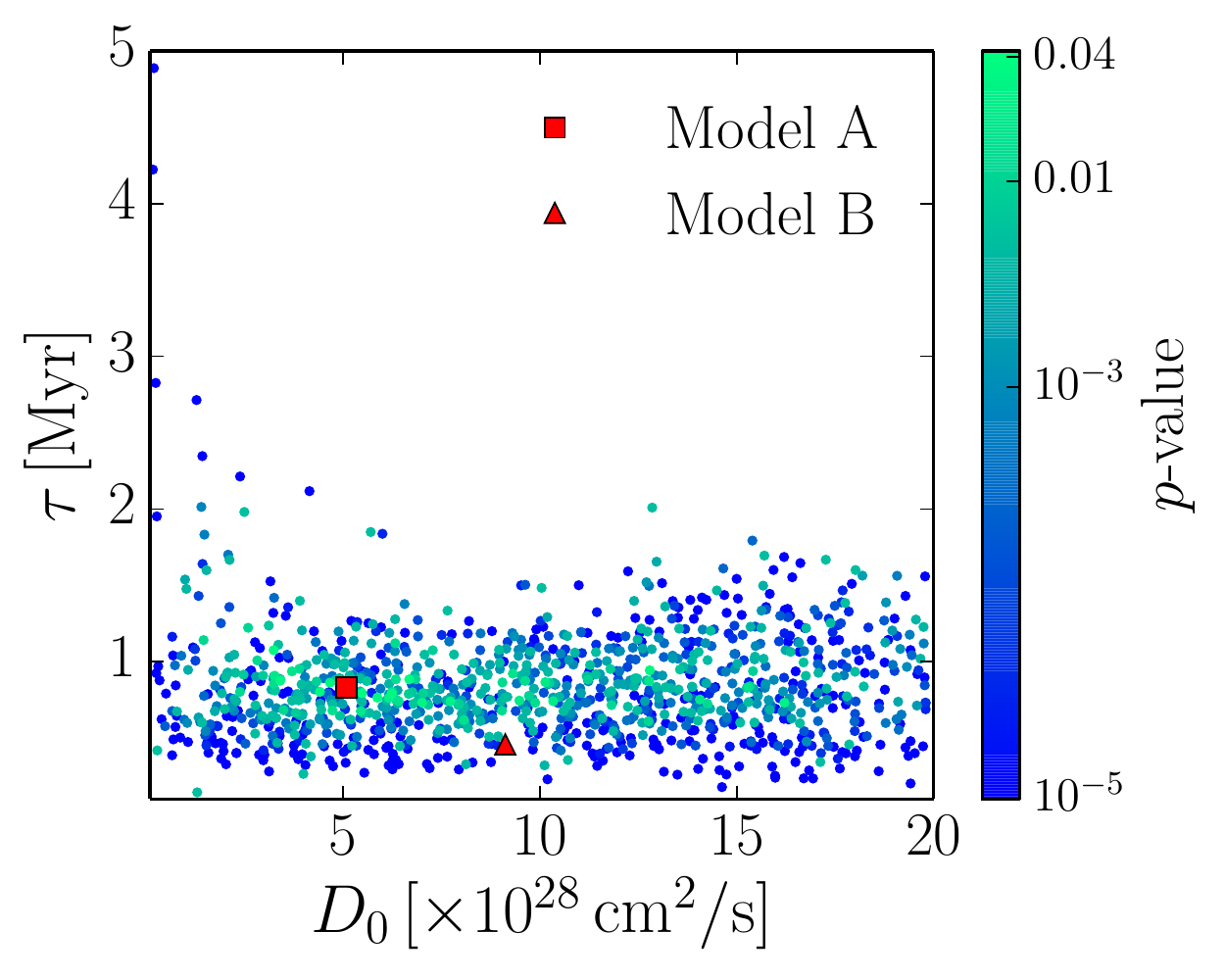}
\includegraphics[width=0.47\linewidth]{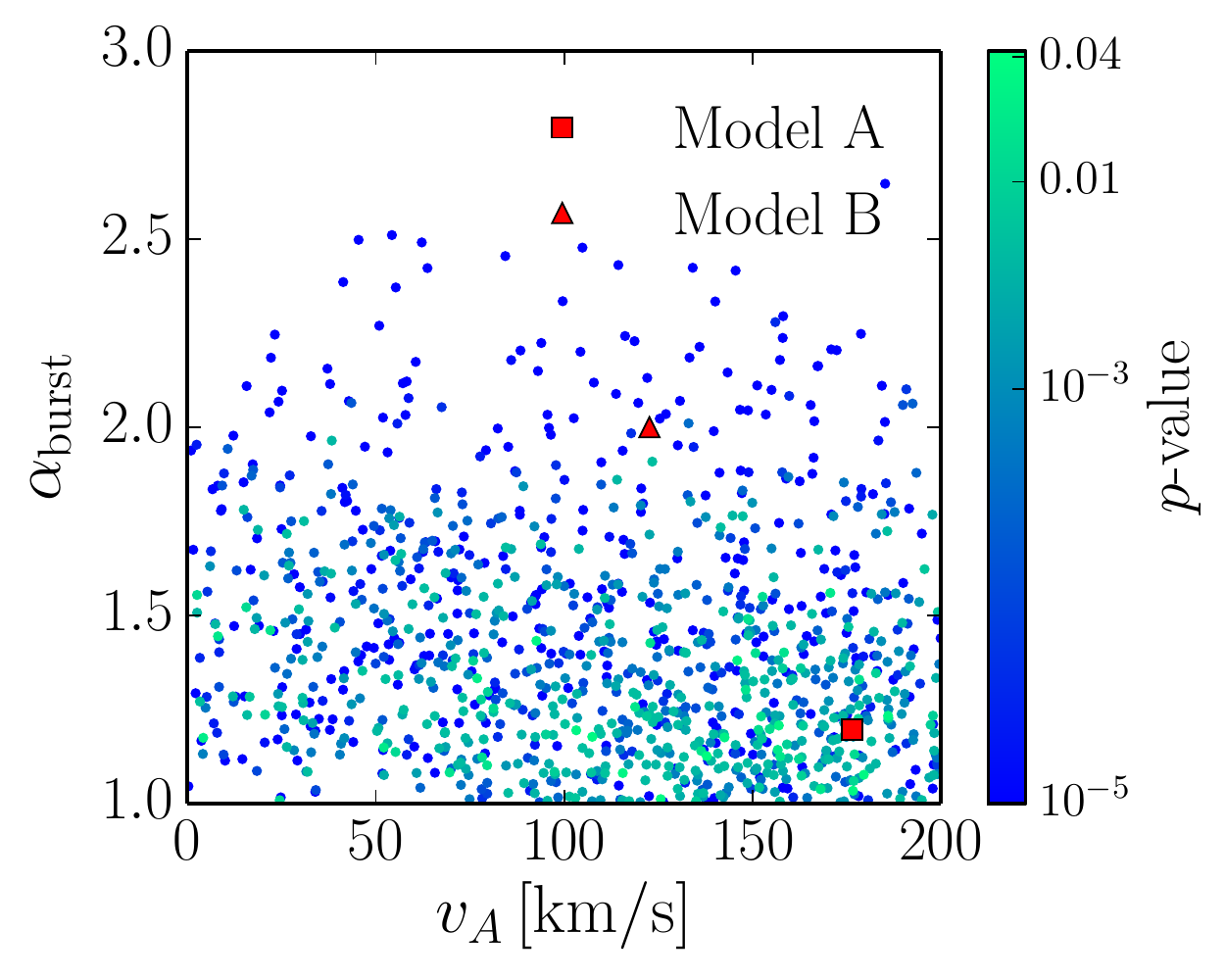}
\includegraphics[width=0.47\linewidth]{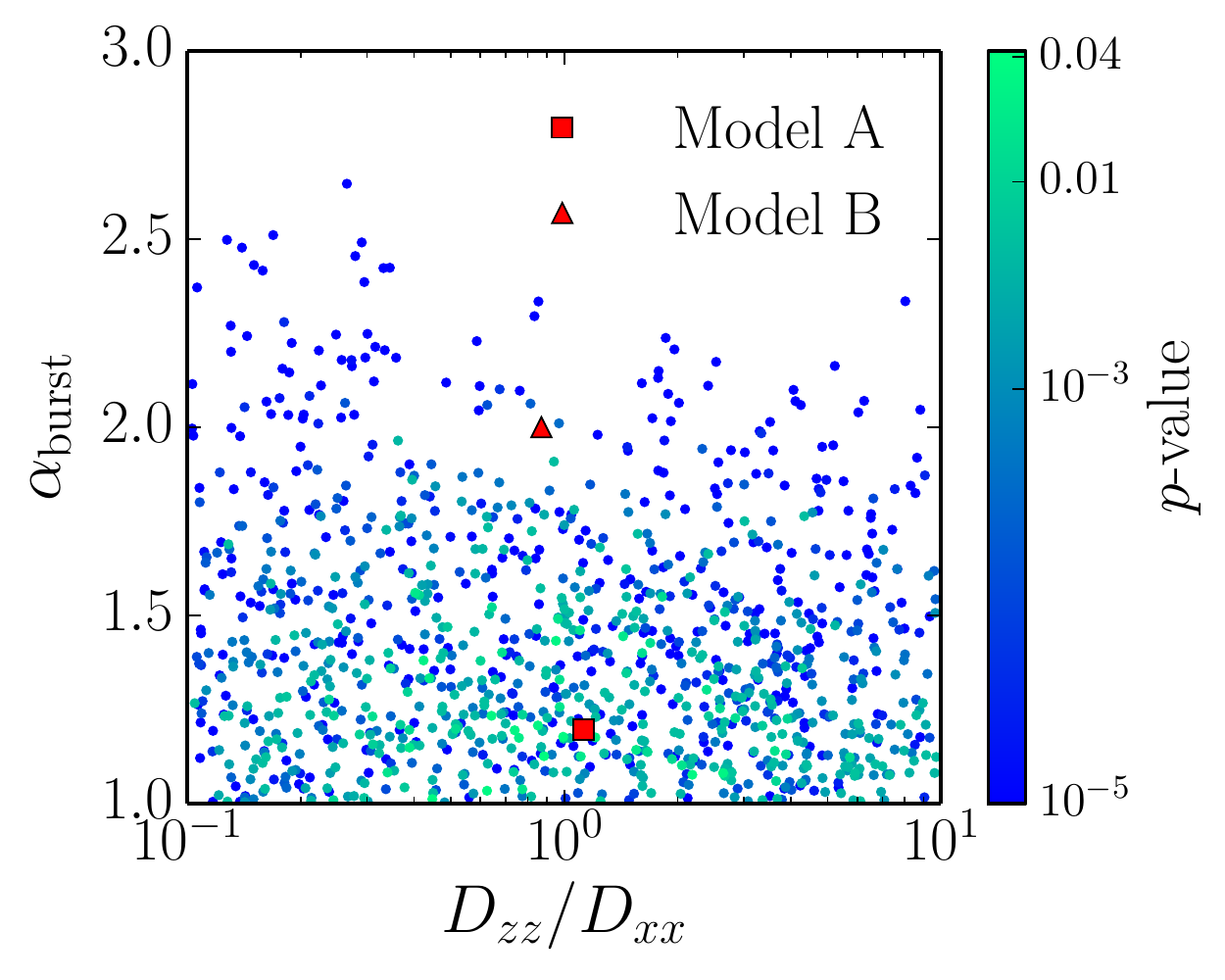}
\end{center}
\caption{The results of our scan over the parameters described in Tab.~\ref{tab:ranges}, for a single outburst.
The GC GeV excess is best-fit by an outburst with an age of about $\tau \sim 1$~Myr, a hard injected spectral index ($\alpha < 1.5$), and with strong re-aceleration ($v_A > 100$~km/s). Anisotropic diffusion does not significantly improve the overall fit.}
\label{fig:contours}
\end{figure}

In figure~\ref{fig:contours}, we show the results of our scan over the parameters described in Tab.~\ref{tab:ranges}.
From the distribution of the models in the $\tau$-$D_0$ plane, we find that the best fits occurs for outbursts with an age of about $\tau \sim 1$~Myr.  In contrast, the fits are not very sensitive to the choice of $D_0$, confirming that the age of the outburst is mainly fixed by the energy loss time-scale (as opposed to the timescale for diffusion).
From the second panel, we see that the spectrum of the excess is best reproduced for hard injection indices ($\alpha < 1.5$) and that strong re-aceleration ($v_A > 100$~km/s) is favored. 
In the third panel, we learn that anisotropic diffusion does not significantly improve the overall fit.

\begin{figure}
    \begin{center}
        \includegraphics[width=0.9\linewidth]{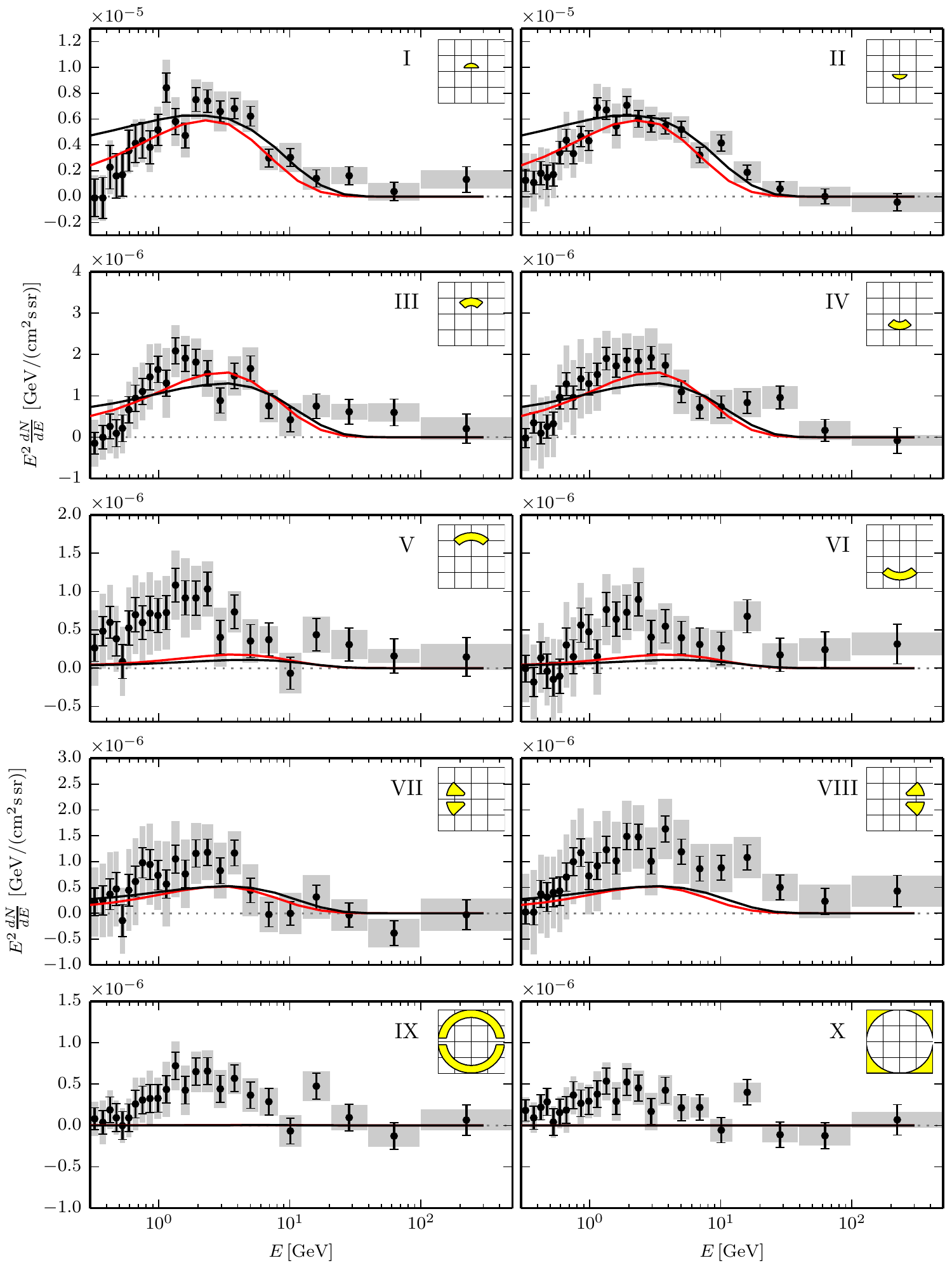}
    \end{center}
    \caption{The gamma-ray spectrum predicted from ten sub-regions of the Inner Galaxy, $40^\circ\times40^\circ$, for the single outburst Models A (red) and B (black); see Tab.~\ref{tab:benchmark}. These results are compared to the observed characteristics of the GC GeV excess, as presented in Ref.~\cite{Calore:2014xka}.  Although neither of these models provide a particularly good fit ($p$-values of 0.04 and 0.0004, respectively), the hard spectral index ($\alpha=1.2$) of Model A allows it to better accommodate the data.}
    \label{fig:benchmark_scan}
\end{figure}

From the scan results, we focus now on the best-fit model, which we denote Model A, and compare these results to the best fit model obtained after bounding  $2 < \alpha < 2.4$ (as considered in Ref.~\cite{Petrovic:2014uda}), which we denote as Model B. The model parameters are given in the first two columns of Tab.~\ref{tab:benchmark}. Note that the normalization of Model A is not at all implausible, being similar to the kinetic energetic released by a few supernova explosions. In figure~\ref{fig:benchmark_scan} we show the spectrum predicted by these two models in the ten sub-regions across the Inner Galaxy.\footnote{We note that the data points that we present are those directly from Ref.~\cite{Calore:2014xka}, 
and remind that within their error bars the data points within each window can move upwards or 
downwards in a correlated fashion based on the covariance matrix.}
Although spectral indices around $\alpha \sim 2$ may be theoretically preferred within the context of Fermi acceleration, we find that much harder spectral indices are favored by the data and that restricting the analysis to $\alpha > 2$ worsens the fit significantly (the $p$-value for Model A is 0.04, whereas it is 0.0004 for Model B).  The hard injected spectral index and high Alfv$\acute{\textrm{e}}$n speed characterizing Model A are responsible for the observed peak in the spectrum of the excess at an energy of a few GeV.

Even these best-fit single outburst models, however, fail to provide a good fit to the observed characteristics of the GC GeV excess, providing $p-$values consistently lower than 0.04. With this in mind, we turn in the following subsection to models with multiple outbursts.

\begin{table}
\begin{tabular}{lccc}
\toprule
Parameter & Model A & Model B & Model C \\
\midrule
$\alpha_1$ & 1.2 & 2.0 & 1.1\\
$\alpha_2$ & NA & NA & 1.0\\
$E_{\rm cut,1}$ &1 TeV & 1 TeV & 20 GeV \\
$E_{\rm cut,2}$ & NA & NA & 60 GeV\\
$\tau_1$ (Myr)    & 0.83 & 0.46 & 0.1\\
$\tau_2$ (Myr)    & NA & NA & 1.0\\
$N_1$ ($10^{51}$ erg) & 2.89 & 9.87  & 0.1\\
$N_2$  ($10^{51}$ erg) & NA & NA  & 0.88\\ 
\hline
$\delta$  & 0.20& 0.23 & 0.3\\
$D_0$ ($10^{28}$ cm$^2$/s)   & 5.08 & 9.12 & 9.0 \\
$D_{zz}/D_{xx}$ & 1.12 & 0.87 & NA\\
$v_A$ (km/s)    & 176 & 122 & 150\\
$B_0$ ($\mu$G)  & 11.5 & 11.5 & 11.7 \\
$r_c$ (kpc)  & 10.0 & 10.0 & 10.0 \\
$z_c$ (kpc)  & 2.0 & 2.0 & 0.5 \\
$dv_c/dz$ (km/s/kpc) &0.0 & 0.0& 0.0\\
ISRF &  1.0, 1.0 & 1.0, 1.0 & 1.8, 0.8 \\
\hline
$\chi^2$ ($p-$value) & 277 (0.04) & 317 (0.0004) & 261 (0.14) \\
\bottomrule
\end{tabular}
\caption{Parameter values for our three main benchmark models. Models A and B include a single cosmic-ray outburst, as described in Sec.~\ref{Sec:single} and shown in figure~\ref{fig:benchmark_scan}. These two sets of parameters represent the best-fits found when the injected spectral index is allowed to float freely (Model A) and when it is constrained to $\alpha \ge2$ (Model B). Models with very hard injected spectral indices, $\alpha \simeq 1.2$, are strongly preferred by the data. Model C includes two outbursts, as described in Sec.~\ref{sec:2bursts}. The two outburst model provides a significantly better fit to the GC GeV excess. The parameters given under the label ``ISRF'' denote the normalizations of the optical and infrared components of the interstellar radiation field, in units relative to the models of Ref.~\cite{Porter:2005qx} for Models A and B, and of Ref.~\cite{GALPROPSite} (\texttt{Galprop v54}) for Model C.}
\label{tab:benchmark}
\end{table}

\subsection{Two outburst model}
\label{sec:2bursts}

As discussed in the previous subsection, a single outburst of cosmic-ray electrons does not appear to be capable of accounting for for the morphological 
characteristics of the GC GeV excess, under-predicting the amount of emission from more than $8^{\circ}$ away from the GC, and from within the innermost $2^{\circ}$ surrounding the GC. This motivates us to consider scenarios with two cosmic-ray electron outbursts, which we take to have occurred 0.1 Myr and 1 Myr ago. Since cosmic-ray electrons tend to propagate outwards from the center of the Galaxy, we expect that the cosmic-ray electrons from the older outburst will be responsible for generating the gamma-ray emission in the outer part of the excess, while the younger outburst will generate most of the excess gamma-ray emission in the inner several degrees. As we will discuss in Sec.~\ref{sec:morphology}, additional (and even more recent) outbursts may be required to accommodate the characteristics of the GeV excess observed from the innermost degree or two surrounding the GC. 


In order to study scenarios featuring two cosmic-ray outbursts, we considered a variety of models, with differing diffusion coefficients, Alfv$\acute{\textrm{e}}$n speeds, convection velocities, interstellar radiation field densities, and magnetic field distributions. The details of this exercise are described in Appendix~\ref{app:35models}. Unlike in the case of models with a single outburst, we have restricted our two outburst models to isotropic diffusion ($D_{xx}=D_{zz}$).

From the wide range of two outburst models we have studied, the best-fit was found for the model that we denote as ``Model C'', whose characteristics are given in the last column of Tab.~\ref{tab:benchmark}. Note that the two outbursts have significantly different values for their cutoff energies in this model, $E_{\rm cut}=20$ GeV and 60 GeV, respectively.  The higher energy cutoff for the older outburst is necessary to ensure that, after accounting for energy losses, the gamma-ray spectrum is approximately uniform over the region of the GeV excess. 
Note that for the single outburst models, as discussed in section~\ref{Sec:single}, the exact value of the cutoff 
doesn't strongly impact the fit, as long as it is $\mathcal{O}(1)$ TeV. This is because of the rapid energy losses experienced by TeV-scale electrons (which scale as $dE_e/dt \propto E_e^2$). For the models considered in this section, it is the combination of the two outbursts that fits the data,
and we find that values of $E_{\rm cut} \simeq 50-100$ GeV for the older outburst yield the best fits.\footnote{In the single outburst models, the fit is most impacted by the regions I and II of the Inner Galaxy, while in the two outburst models the spectral properties of the 1 Myr old outburst are most constrained by regions III-VIII, which prefer lower-energy cutoffs.}

\begin{figure}
    \begin{center}
        \includegraphics[width=0.8\linewidth]{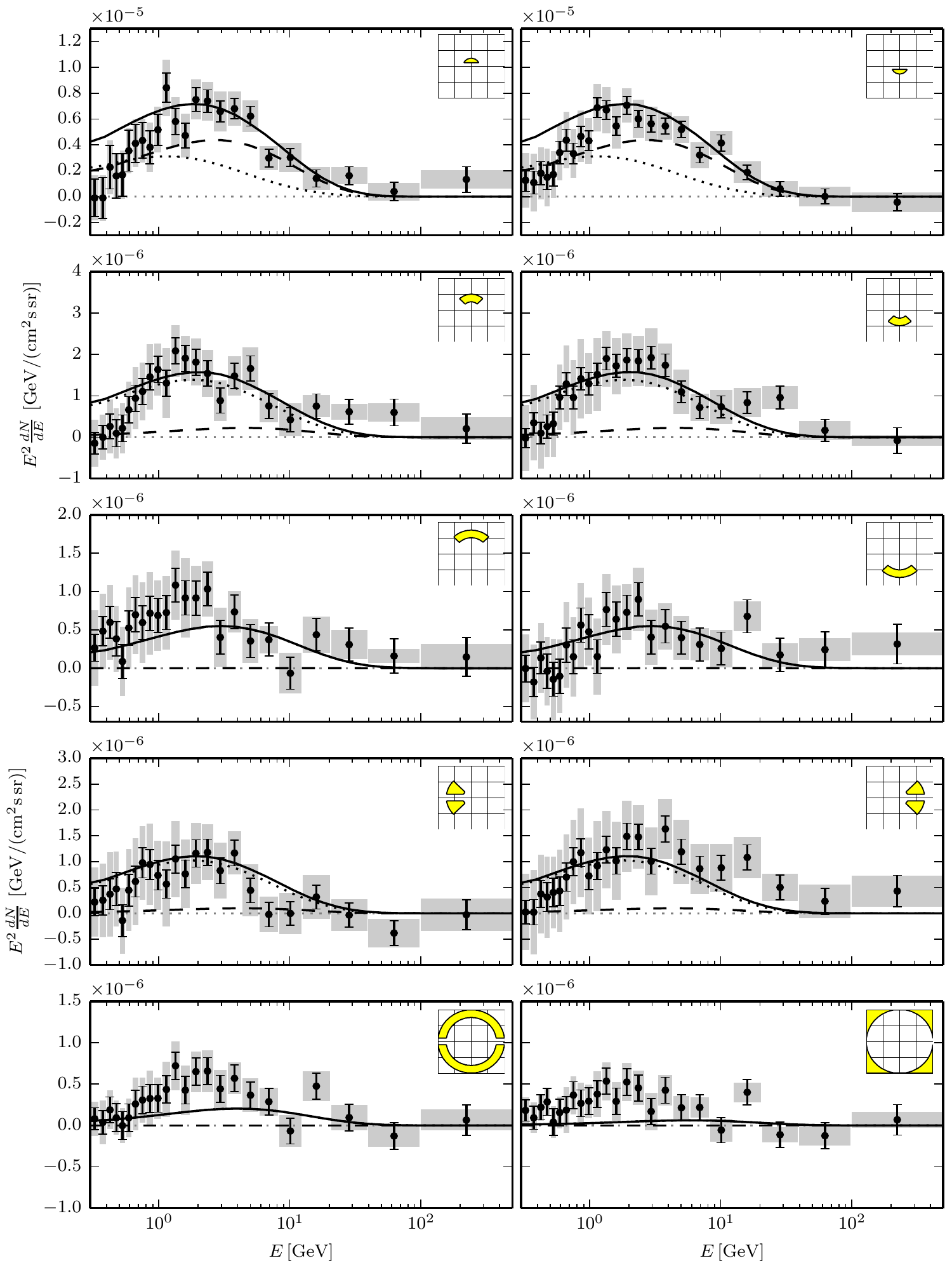}
    \end{center}
    \caption{As in figure~\ref{fig:benchmark_scan}, but for the case of a model with two cosmic-ray outbursts (Model C, see Tab.~\ref{tab:benchmark}). The dashed (dotted) line denotes the contribution from the younger (older) outburst, and the solid line represents the sum of their contributions. This model provides the best fit of all those we have considered in this study, yielding a $p$-value of 0.14.}
    \label{fig:C20B_best_GC}
\end{figure}

In figure~\ref{fig:C20B_best_GC}, we show the gamma-ray emission from the combination of these two outbursts in the 10 sub-regions of the GC GeV excess \cite{Calore:2014xka}.
The best-fit model yields $\chi^2 = 261$, corresponding to a $p$-value of 0.14.
As one can see, the flux from the younger outburst (dashed line) produces most of the emission 
at $2^{\circ}-5^{\circ}$ from the GC, while the 1 Myr old outburst accounts for the majority of the excess emission at larger angles.

A few comments regarding the parameter choices of Model C are in order. In particular, the high value of the Alfv$\acute{\textrm{e}}$n speed, $v_{A} = 150$ km s$^{-1}$, may strike some readers as unconventional. Although local cosmic-ray measurements imply values for this quantity that are on the order of a few tens of km/s, the conditions of the Inner Galaxy are not well understood or constrained by observations. For this reason, we remain highly agnostic as to the values of this and other propagation parameters (see also Appendix~\ref{app:35models}). Another notable characteristic of Model C is its very hard spectral indices, $\alpha=1.1$, 1.0, and spectral cutoffs at 20 and 60 GeV. We find it plausible that such spectra might result under the conditions within the inner $\mathcal{O}(10)$ pc, where diffusive re-acceleration plays an important role (hardening the spectral index) and where there is strong turbulence and large amounts of energy in the magnetic field (causing the spectral cutoff from synchrotron energy losses). Finally, we note that we require $\sim10^{50}-10^{51}$ ergs in cosmic-ray electrons to fit the GeV excess, similar to the energetics required to generate the \textit{Fermi} Bubbles.

\begin{figure}
    \begin{center}
        \includegraphics[width=0.8\linewidth]{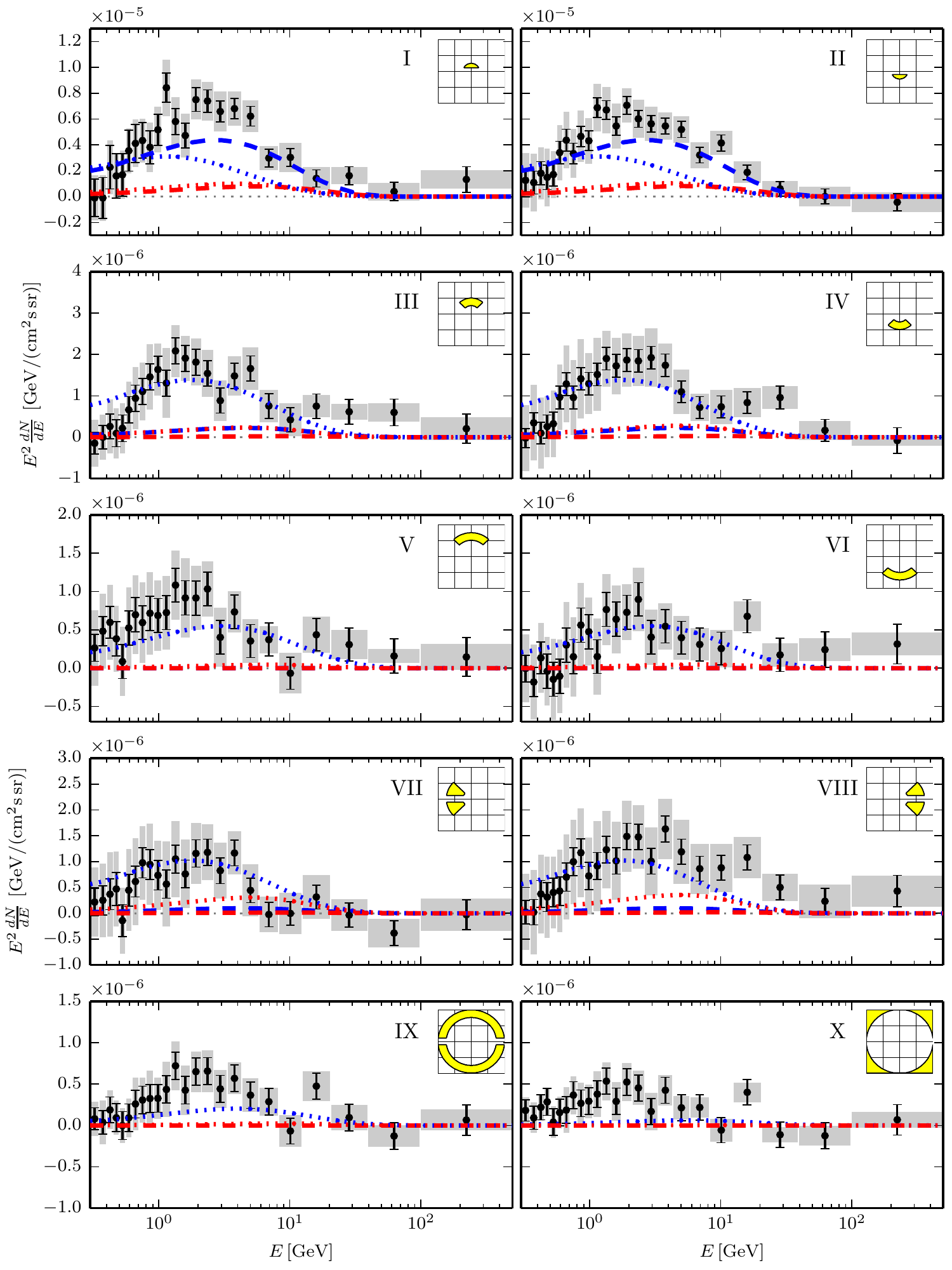}
    \end{center}
    \caption{As in figure~\ref{fig:C20B_best_GC}, but showing separately the contributions from inverse Compton scattering (blue) and Bremsstrahlung emission (red) for our best-fit two outburst model (Model C, see Tab.~\ref{tab:benchmark}). The Bremsstrahlung component has been multiplied by a factor of 100 in each frame in order to be distinguishable from zero. Once again, the dashed and dotted lines denote the contributions from the younger and older outburst, respectively.}
    \label{fig:C20B_ICS_and_Bremss}
\end{figure}

In figure~\ref{fig:C20B_ICS_and_Bremss}, we show the contributions in Model C from ICS (blue) and Bremsstrahlung (red), from the 0.1 Myr old (dashed) and the 1 Myr old (dotted) outbursts. The Bremsstrah-lung emission (which has been multiplied by a factor of 100 in order to be distinguishable from zero) is always subdominant to the ICS emission in this model (except in the region near the Galactic Plane, as discussed in Sec.~\ref{sec:morphology}).

\subsection{The morphology of the gamma-ray emission from the Galactic Center and Galactic Plane}
\label{sec:morphology}

In this section, we discuss aspects of the morphology of the gamma-ray emission from cosmic-ray outbursts that is not directly evident from the plots shown in the previous subsections, focusing on the regions that lie within $2^{\circ}$ of the GC or along the Galactic Plane.  

\subsubsection{The Galactic Center}
\label{gcsection}

\begin{figure}
    \begin{center}
        \includegraphics[width=0.8\linewidth]{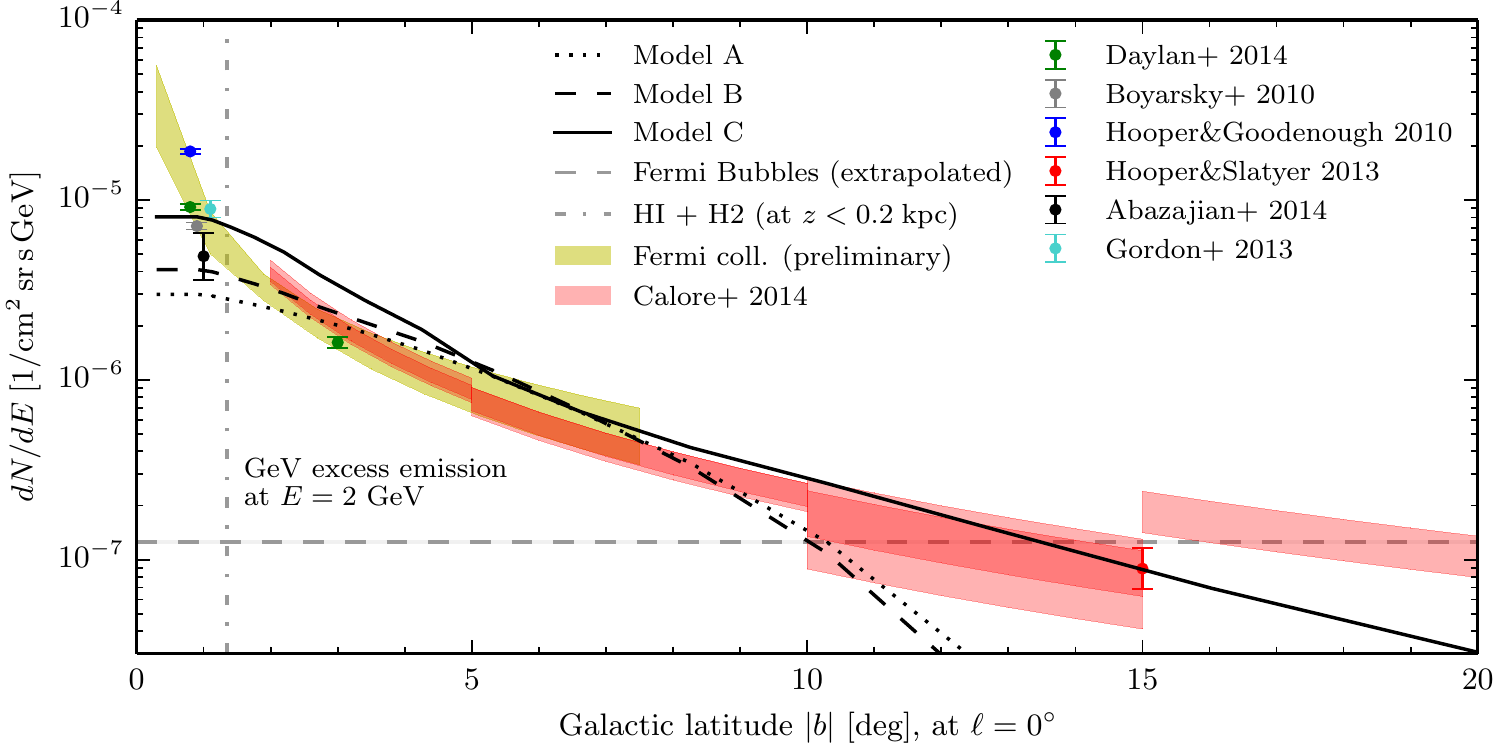}
    \end{center}
    \caption{The angular profile of the GC excess flux, as evaluated at 2 GeV, for single outburst Models A and B, and for the two outburst Model C (see Tab.~\ref{tab:benchmark}). Single outburst models can fit the morphology of the excess in the range between $\sim 2^{\circ}-8^{\circ}$ away from the GC, while two outburst 
    models can fit the excess morphology between $\sim 1^{\circ}-15^{\circ}$. Additional recent outbursts, and/or a population of centrally located millisecond pulsars, are required in order to account for the emission observed from the innermost degree surrounding the GC.  For details on the observational data, see Ref.~\cite{Calore:2014nla} and Ref.~\cite{simonaTalk}.}
    \label{fig:profile}
\end{figure}

In figure~\ref{fig:profile}, we show the angular profile of the gamma-ray emission predicted by the cosmic-ray outburst Models A, B and C. Models with a single outburst (Models A and B), can only fit the data between approximately $2^{\circ}$ and $8^{\circ}$, while the two outburst model (Model C) 
can accommodate the observed morphology of the GC GeV excess between approximately $1^{\circ}$ and $15^{\circ}$ from the GC. The main difficulty in accommodating the observed morphology of the excess with cosmic-ray outbursts is that the data favor a concave profile, while individual outbursts produce a convex distribution. The sum of the contributions from multiple carefully placed convex-profiled outbursts can, however, resemble the observed concave distribution. 

At angles greater that $15^{\circ}$ from the GC, the excess signal is faint (and is difficult to distinguish from the  \textit{Fermi} Bubbles) and has relatively little statistical impact on the fits. Of greater importance is the morphology of the excess within the innermost $1^{\circ}$, which is measured to continue rising to at least within $\simeq 0.05^{\circ}$ (or $\simeq 7.4$ pc) of the GC~\cite{Daylan:2014rsa}. None of the models considered in this study can accommodate this observed feature of the GC GeV excess.

One possible solution to this discrepancy would be to introduce additional outbursts, which took place in the more recent past. Such outbursts could dominate the emission within the innermost $1^{\circ}$ without significantly impacting the emission from much further away from the GC (similar to how the 0.1 Myr old outburst does not contribute very much to the emission further away than $5^{\circ}$). In order for the flux of the gamma-ray excess to continue increasing to within $0.05^{\circ}$ of the GC, the most recent outburst(s) must have taken place very recently, within the past several hundred years.

To explain the rising gamma-ray flux at $\simeq 0.05^{\circ}$ ($\simeq 7$ pc) from the GC~\cite{Daylan:2014rsa} in an outburst scenario (or any other scenario relying on emission from ICS), there must exist a highly cusped density of cosmic-ray electrons within the tens of parsecs immediately surrounding the GC. These electrons will invariably generate a radio flux via synchrotron emission, which can be constrained by observations. The authors of Refs.~\cite{Bringmann:2014lpa} and~\cite{Cholis:2014fja} each studied this constraint within the context of DM annihilations being responsible for the GC GeV excess, making somewhat different assumptions regarding the magnetic and radiation fields present within the innermost parsecs of the Milky Way. In either case, these constraints on annihilating DM can be translated to apply to the case of cosmic-ray outbursts by the following procedure.  For cosmic rays generating gamma rays through ICS, the ratio of power in radio emission to that in gamma rays is simply given by the ratio of the energy densities in magnetic and radiation fields:
\begin{equation}
\frac{F_{\rm radio}}{F_{\gamma}} \bigg|_{\rm CR}= \frac{\rho_B}{\rho_{\rm rad}},
\end{equation}
whereas DM annihilations to prompt photons leads to a lower ratio:
\begin{eqnarray}
\frac{F_{\rm radio}}{F_{\gamma}} \bigg|_{\rm DM} &=& \frac{B_e \bigg(\frac{\rho_B}{\rho_B+\rho_{\rm rad}}\bigg)}{B_e \bigg(\frac{\rho_{\rm rad}}{\rho_B+\rho_B}\bigg) +B_{\gamma}}, 
\end{eqnarray}
where $B_e$ and $B_{\gamma}$ are the fractions of the energy generated in DM annihilations that go into electrons/positrons and prompt photons, respectively.  If we keep the energy output in gamma rays (from prompt annihilations and/or ICS) fixed in these two scenarios, we find the following ratio of power in synchrotron emission:
\begin{eqnarray}
\frac{F_{\rm radio, CR}}{F_{\rm radio, DM}}&=& 1 + \frac{B_{\gamma}}{B_e} \bigg(\frac{\rho_B+\rho_{\rm rad}}{\rho_{\rm rad}}\bigg).
\end{eqnarray}
For the conservative case of $\rho_B \ll \rho_{\rm rad}$, and for annihilations to $b\bar{b}$, this ratio is approximately 2.5, and therefore radio constraints will be at least a factor of 2.5 times more stringent in the cosmic-ray outburst case than in the case of annihilating DM. If we consider the more conservative ({\it i.e.} less restrictive) assumptions of Ref.~\cite{Cholis:2014fja}, the constraint from radio observations of the GC requires that the profile of the electron distribution must become flat before reaching the innermost $r\sim1$ pc, which is marginally consistent with the morphology of the excess. For the less conservative assumptions made in Ref.~\cite{Bringmann:2014lpa}, the derived radio constraints are not compatible with a cosmic-ray outburst interpretation. Taken together, these results indicate that radio observations are compatible with a cosmic-ray outburst interpretation of the central GeV excess only if high radiation field densities and strong convective winds are present in the innermost parsecs of the Galaxy (such as considered in Ref.~\cite{Cholis:2014fja}).

In light of the apparent difficulties for cosmic-ray outburst scenarios to explain the entirety of the excess emission observed from the innermost degree around the GC, one could also consider a hybrid scenario, in which two or more cosmic-ray outbursts generate the excess beyond $1^{\circ}$--$2^{\circ}$, while a population of centrally located unresolved millisecond pulsars is responsible for the majority of the emission from the innermost volume around the GC. Millisecond pulsars have often been discussed as a possible source of the GC GeV excess~\cite{Hooper:2010mq,Abazajian:2012pn, Abazajian:2014fta,Cholis:2014lta,Calore:2014oga,Hooper:2013nhl,Cholis:2014noa,Petrovic:2014xra}, motivated by the fact that the gamma-ray spectra measured from this class of objects is similar to that of the observed excess.  And while it has been argued that the lack of pulsar-like point sources detected and resolved by \textit{Fermi} in the regions north and south of the GC significantly disfavor the possibility that millisecond pulsars could account for the entirety of the GeV excess~\cite{Cholis:2014lta} (see also Refs.~\cite{Calore:2014nla, Petrovic:2014xra}), bright backgrounds in the innermost degree around the GC make \textit{Fermi} much less sensitive to point sources in this region of the sky.  To account for half of the GeV excess observed from within in the innermost degree (the other half coming from cosmic-ray outbursts), would require approximately 225 millisecond pulsars, of which we expect only $\sim$\,7 to be quite bright, $L_{\gamma} > 10^{35}$ erg/s (integrated above 0.1 GeV). Observations from \textit{Fermi} cannot, at this time, rule out the existence of a centrally concentrated millisecond pulsar population of this size.

One argument that can be made against the existence of a large centrally concentrated population of millisecond pulsars makes use of the connection between millisecond pulsars and low-mass X-ray binaries (LMXBs)~\cite{Cholis:2014lta}. Most millisecond pulsars are thought to have evolved from LMXBs, and the abundance of LMXBs in a region is expected to be highly correlated to the number of millisecond pulsars that are present. Focusing on the brightest ($L>10^{36}$ erg/s) LMXBs, for which we are believed to have a complete inventory, we can compare the number of bright LMXBs in globular clusters to the gamma-ray emission from such systems (assumed to be produced by millisecond pulsars), and compare this ratio to that found in the GC. If half of the GeV excess from the innermost degree around the GC was produced by millisecond pulsars, this exercise predicts that approximately 25 bright LMXBs should also be present in the same region. In contrast, \textit{INTEGRAL} (which has sensitivity in the direction of the GC well above the level required to detect such bright sources) has detected only one bright LMXB candidates in this region of the sky~\cite{Revnivtsev:2008fe}, suggesting that relatively little of the GeV excess originates from MSPs.

\subsubsection{Bremsstrahlung emission along the Galactic Plane}
\label{gpsection}

\begin{figure}
    \begin{center}
        \includegraphics[width=0.48\linewidth]{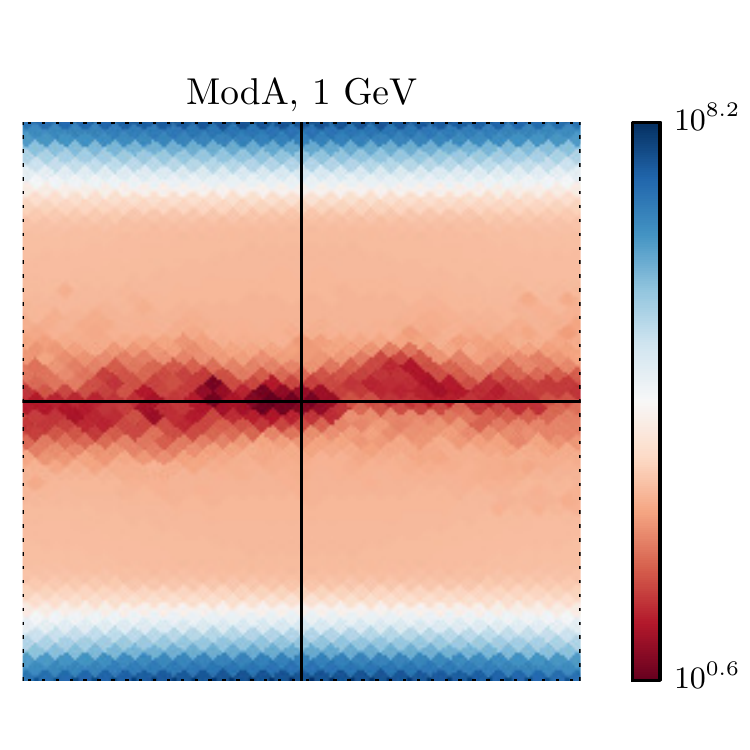}
        \includegraphics[width=0.48\linewidth]{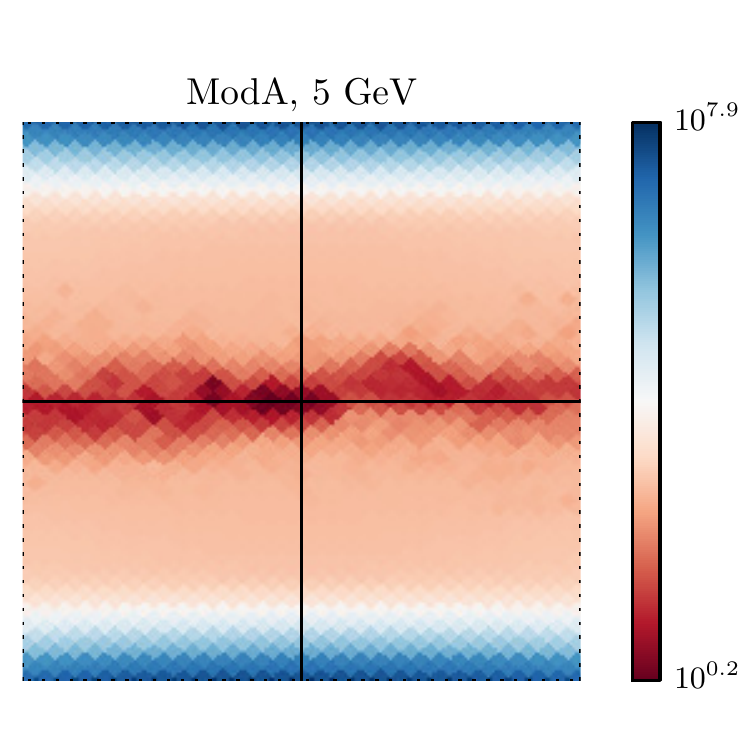} \\
        \includegraphics[width=0.48\linewidth]{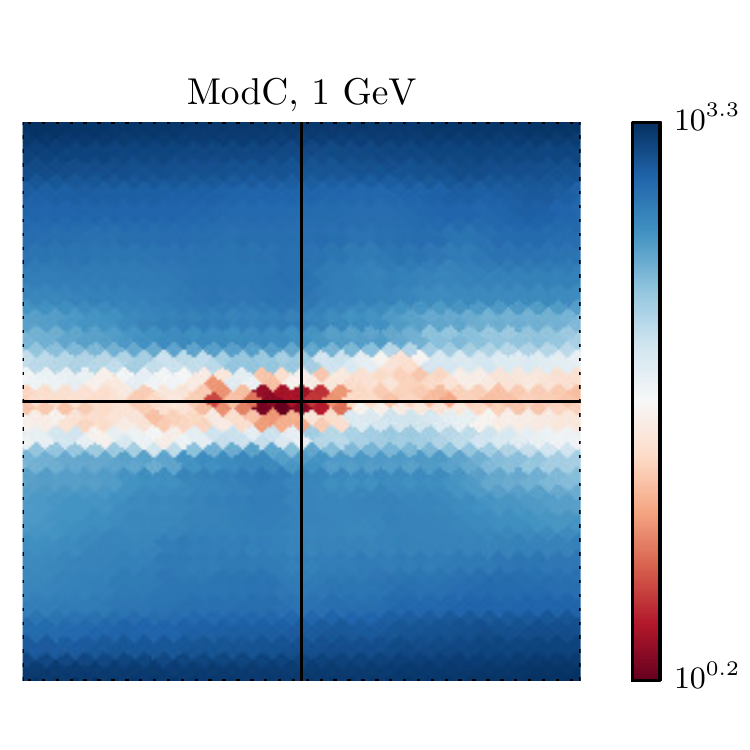}
        \includegraphics[width=0.48\linewidth]{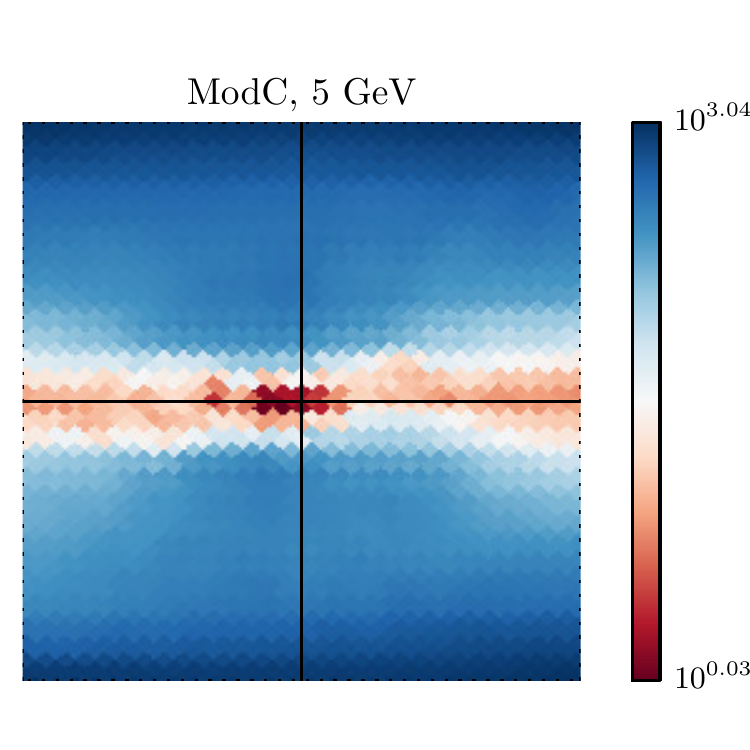}
    \end{center}
    \caption{The ICS-to-Bremsstrahlung ratio map for the single outburst Model A (upper frames) and the two outburst Model C (lower frames); see Tab.~\ref{tab:benchmark}. The maps cover the $20^{\circ} \times 20^{\circ}$ region centered at the GC, without masking the 
   Galactic Plane. The color-bar is in log$_{10}$ scaling. Even in the pixels closest to the GC, 
   the ICS-to-Bremsstrahlung ratio is always greater than unity, although the Bremsstrahlung component can be as bright as $\sim$\,$30-100\%$ of that from ICS within the innermost $1^{\circ}$, and $\sim$\,$10\%$ elsewhere along the Galactic Plane.}
    \label{fig:ICStoBremssMap}
\end{figure}

In figure~\ref{fig:C20B_ICS_and_Bremss}, the gamma-ray emission is dominated by ICS over each of the regions of the sky shown, with Bremsstrahlung playing only a very subdominant role. Due to the nearly homogeneous nature of the interstellar radiation field throughout the Inner Galaxy, the emission from ICS is predicted to be approximately spherically symmetric with respect to the GC, in agreement with the observed morphology of the GeV excess. In the regions near the Galactic Plane, however, the Bremsstrahlung component is not necessarily negligible, as the gas densities are much higher within the volume of the Galactic Disk. To explore this feature, we plot in figure~\ref{fig:ICStoBremssMap} the ratio of ICS-to-Bremsstrahlung emission from either a single outburst (model A) or from a combination of two outbursts (model C), throughout the $20^{\circ} \times 20^{\circ}$ box around the GC, including the Galactic Disk where 
the gas density peaks. In the region within $1^{\circ}$ of the GC, the intensity of the Bremsstrahlung emission can be as large as 30-100\% of the ICS component. Excluding the window around the GC, the region along the Galactic Plane can exhibit Bremsstrahlung-to-ICS ratios on the order of $\sim$10\%.\footnote{In figure~\ref{fig:ICStoBremssMap}, we show ratio maps as evaluated at both 1 and 5 GeV, in order to test for any energy dependance. The similarity of these frames indicates that any such variation with energy is mild.} This will lead to a gamma-ray morphology that is slightly elongated along the direction of the Galactic Plane. In Sec.~IV of Ref.~\cite{Daylan:2014rsa}, constraints were placed on such departures from spherical symmetry, finding that axis ratios which depart by more than $\sim$20\% from unity along the Galactic Plane are strongly disfavored. Although this constraint is in mild tension with the results shown in figure~\ref{fig:C20B_ICS_and_Bremss}, it does not significantly disfavor the models under consideration in this study.  If future observations were to further tighten this constraint, this could be used to test cosmic ray outburst models of the GC excess.

\section{Discussion and Conclusion}
\label{sec:conclusions}

In this paper, we have presented a detailed study of the gamma-ray emission from
leptonic cosmic-ray outbursts from the Galactic Center, and considered this as a possible source for the GeV excess observed by the \textit{Fermi}-LAT.  We solved the full cosmic-ray propagation equations numerically, allowing us to fully take into account
spatially dependent energy losses, variations in the cosmic-ray propagation
conditions, and effects including re-acceleration and anisotropic diffusion, all of
which had been neglected previously.  We performed a systematic scan over source and
propagation parameters using nested sampling techniques, and fitted the results to the spectrum and morphology of the GeV excess over the region of the Inner Galaxy.

For models with a single outburst, as had been considered previously in the literature, we find that it is not possible to explain the overall morphology of the observed GeV excess. In particular, despite allowing many parameters to vary freely, we find that single outburst models uniformly underproduce the amount of excess emission that is observed from the innermost regions, outermost regions, or both, throughout the Inner Galaxy. The best-fit models include very strong re-acceleration (Alfv$\acute{\textrm{e}}$n speeds of around 175 km/s) and require electrons to be injected from the Galactic Center with an extremely hard spectral index, around $\alpha \simeq 1.2$. Dramatically worse fits were found for models with a softer spectral index, as are generally predicted from Fermi acceleration.

Motivated by the inability of single outburst models to accommodate the observed characteristics of the GeV excess, we also explored models that included two leptonic outbursts. Focusing on models with an older (1 Myr) and younger (0.1 Myr) outburst, we find that such scenarios could indeed provide much better fits to the data, provided that the injection parameters of both bursts are adjusted
appropriately.  The older outburst in this scenario accounts for the emission at high Galactic latitudes, whereas the younger outburst dominates the excess emission at lower latitudes and closer to the Galactic Center.  If the injected spectra of the two outbursts are tuned accordingly, it is possible that their spectra will be similar after accounting for energy losses that take place during propagation, enabling the gamma-ray emission to exhibit approximately the same spectral shape at all relevant latitudes, consistent with the observed characteristics of the GeV excess.

Our best-fit was found for a model with two outbursts, yielding a fit to the data with a $p$-value of 0.14, which is only modestly lower than those values previously found for annihilating dark matter models ($p \simeq 0.7$ for annihilations to hadronic final states).\footnote{In Ref.~\cite{Calore:2014nla}, $p$-values of 0.35 and 0.37 are quoted for dark matter annihilations to $b\bar{b}$ and $c\bar{c}$, respectively. These values refer to the fitting of the spectrum over the entire $|l|<20^{\circ}$, $2^{\circ}<|b|<20^{\circ}$ region, while the $p$-values quoted throughout this paper are derived from fitting the combination of the ten sub-regions, as shown in Figs.~\ref{fig:benchmark_scan}-\ref{fig:C20B_ICS_and_Bremss}. When this procedure is followed for the same dark matter models, $p$-values of approximately 0.7 are found.}
This model features injected spectral indices of around 1.1 and 1.0, and spectral cutoffs around 20 and 60 GeV for the younger and older outburst, respectively.  
The extremely hard spectral indices necessary to produce the GC
spectrum appear to be incompatible with models of first-order Fermi
acceleration, as well as with the observed electron injection spectra
observed from gamma-ray blazars~\cite{Boettcher:2013wxa}. However, these
spectral indices could plausibly result from the strong diffusive
re-acceleration and turbulence in the region surrounding the Galactic
Center.
The total energy required in cosmic-ray electrons from the younger and older outbursts are $1\times 10^{50}$ ergs and $8.8 \times 10^{50}$ ergs, respectively.   If we take their durations to be $\sim10^4$ years, for example, this implies cosmic ray electron luminosities on the order of $L \sim 10^{38}$ erg/s and $L \sim 10^{39}$ erg/s, below the Eddington limit of Sgr A*~\cite{Cramphorn:2001ye}.

As discussed in sections~\ref{sec:intro} and~\ref{sec:outbursts}, cosmic ray outbursts might originate from 
past activity of the central massive black hole or from a series of starburst events. Some hints suggest that these processes may have occurred on time scales of Myrs ago. As for the massive black hole, its activity is firmly constrained over the past $\sim$10$^3$ years, while less robust constraints on the luminosity of Sgr A* exist for timescales of $10^3 -  10^5$ years~\cite{Ponti:2012pn}.

One other finding of this study is that, even in models with two outbursts, it is not possible to generate the observed morphology of the excess emission from the innermost degree or two around the Galactic Center. To address this, the two outburst model could be augmented with either additional recent outbursts, or with a centrally concentrated population of hundreds of unresolved millisecond pulsars. Although there exist constraints that challenge the viability of each of these scenarios, neither can be entirely excluded at this time.

\textit{In summary}, we find that scenarios involving a series of leptonic cosmic-ray outbursts from the Galactic Center (perhaps augmented by a centrally concentrated population of millisecond pulsars) could potentially generate gamma-ray emission with a spectrum and morphology that is able to account for the GeV excess observed by the \textit{Fermi} gamma-ray telescope.  The approximately uniform spectral shape exhibited by the GeV excess is not generally expected from scenarios featuring multiple cosmic ray outbursts, however, and this feature of the observed excess can be accommodated only if the parameters of the model are selected rather carefully. If variations in the spectrum of the GeV excess across the Inner Galaxy were to be observed in the future, it would provide support for interpretations involving a series of cosmic-ray outbursts.  Similarly, multiple outburst models do not generally produce a gamma-ray signal with radial morphology that follows a power-law (as e.g.~predicted by dark matter annihilation).  The robust characterization of the latitude profile of the observed excess is thus crucial to discriminate between different possible origins.  
 
Although a series of leptonic outbursts as an explanation for the full GeV excess cannot be ruled out observationally, none of the characteristic features of such bursts have been observed up to now.  In contrast, the required injection and propagation parameters need to take extreme values and to be finely adjusted to reproduce all observational aspects of the excess, making this scenario observational viable, but unlikely.

\bigskip
\bigskip

\textbf{Acknowledgments.} We would like to thank John Beacom, Gianfranco Bertone, Sera Markoff and Andrew Taylor for fruitful discussions. 
IC is supported by the US Department of Energy, and would like to thank the Korean Institute for Advanced Study (KIAS) for their hospitality during the progression of this work. CE acknowledges support from the ``Helmholtz Alliance for Astroparticle Physics HA'', funded by the Initiative and Networking Fund of the Helmholtz Association. FC is supported by the European Research Council through the ERC starting grant WIMPs Kairos, P.I. G.~Bertone. TL is supported by the National Aeronautics and Space Administration through Einstein Postdoctoral Fellowship Award No. PF3-140110. CW is P.I.~of the VIDI research programme ``Probing the Genesis of Dark Matter'', which is financed by the Netherlands Organisation for Scientific Research (NWO).  DH is supported by the US Department of Energy under contract DE-FG02-13ER41958. Fermilab is operated by Fermi Research Alliance, LLC, under Contract No. DE- AC02-07CH11359 with the US Department of Energy.  This work has made use of SciPy~\cite{SciPy}, PyFITS\footnote{\url{http://www.stsci.edu/resources/software_hardware/pyfits}}, PyMinuit\footnote{\url{http://code.google.com/p/pyminuit}}, IPython~\cite{IPython}, and HEALPix \cite{Gorski:2004by}. We acknowledge the University of Chicago Research Computing Center for providing support for this work.



\clearpage
\appendix

\section{Description of the two outburst models considered in this study}
\label{app:35models}

In this appendix, we describe the different parameter values we have considered regarding cosmic-ray diffusion, convection, diffusive re-acceleration, 
magnetic fields, and the interstellar radiation field. As the conditions within the Inner Galaxy need not resemble those of our local environment (which are constrained by measurements of cosmic-ray secondary-to-primary ratios), we consider models that span a rather wide range of assumptions.  

Considering many sets of values for the injection and propagation parameters allows us to accomplish two things. First, this procedure allows us to identify the range of models that provides the best fit to the data (this is how we identified Model C, as discussed in the main text). Second, it helps us to address the question of how generic various features are for the gamma-ray emission from cosmic-ray outbursts.

\begin{figure}
    \begin{center}
        \includegraphics[width=0.8\linewidth]{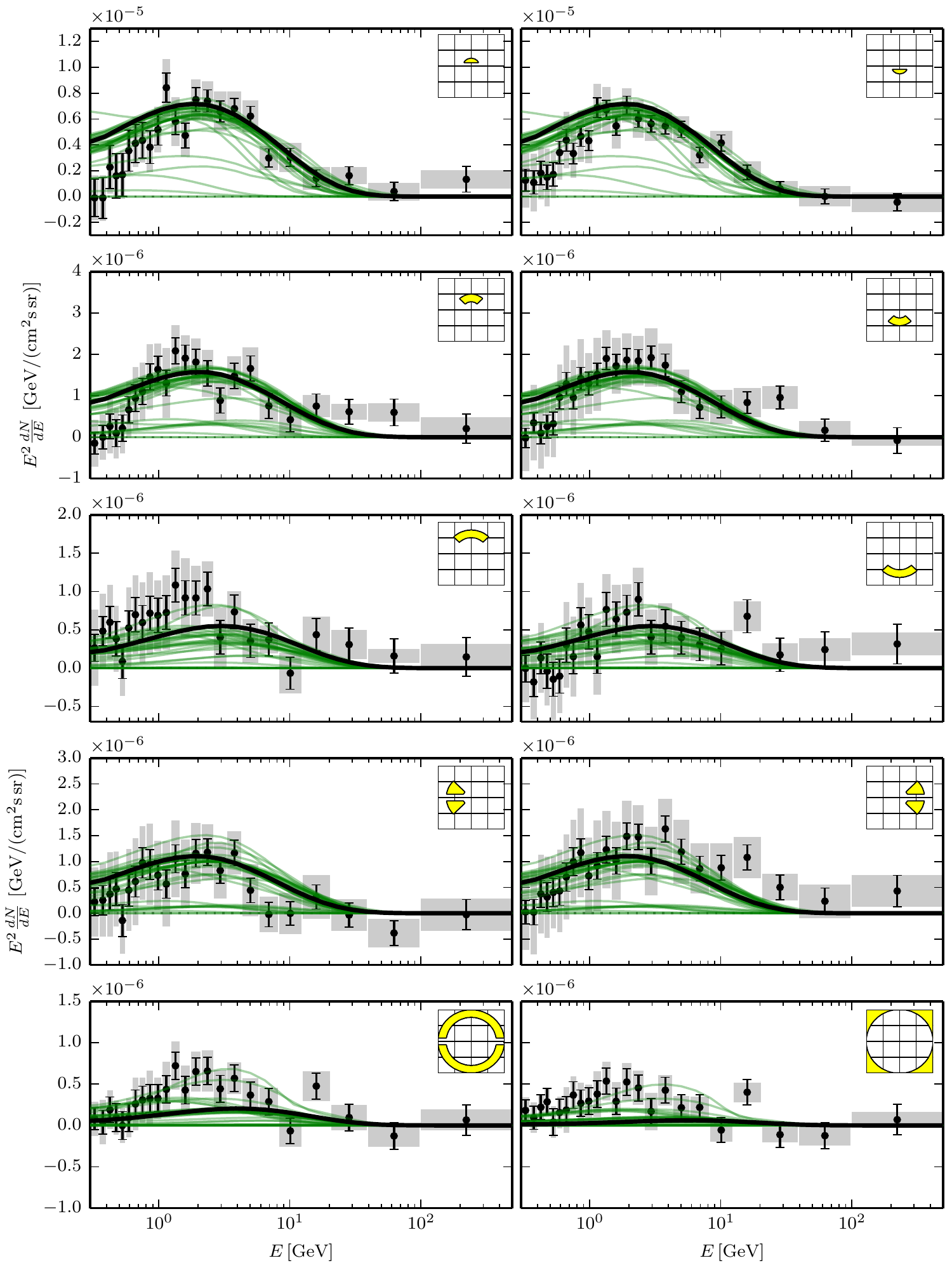}
    \end{center}
    \caption{As in Figs.~\ref{fig:benchmark_scan} and~\ref{fig:C20B_best_GC}, we show the gamma-ray spectrum predicted from ten sub-regions of the Inner Galaxy, for models with two cosmic ray outbursts, using the 35 propagation models described in table~\ref{tab:35Models}. In each case, the injected spectra and normalizations were selected to provide the best fit to the data (as presented in Ref.~\cite{Calore:2014xka}). For each of the 35 propagation models, we show the one combination (individual green lines) that provides the best-fit. The thick black line denotes the overall best-fit model (Model C/XX).}.
    \label{fig:2bursts_variations}
\end{figure}

In figure~\ref{fig:2bursts_variations}, we show the gamma-ray emission predicted in models with two cosmic-ray electron outbursts (a 0.1 Myr and a 1 Myr old), for 35 different propagation models, each described in table~\ref{tab:35Models}. For each of these propagation models, we have allowed the injection parameters to float freely to their best-fit parameters, exploring approximately 3000 parameter sets in total. From this figure, we see that while many of these propagation models dramatically fail to accommodate the observed features of the GeV excess, there exist several that provide a reasonably good fit to the data. This is confirmed in the last column of table~\ref{tab:35Models}, where we list the $p$-values for each best-fit two outburst model.

\begin{table}[t!]
    \centering
    \small
    \begin{tabular}{cccccccccc}
     \toprule
        Model & $D_{0}$ & $\delta$ & $v_{A}$ & $dv_{c}/dz$ & $B_{0}$ & $r_{c}$  & $z_{c}$ & ISRF & $p^{\rm best}$ \\
          Name         & {\footnotesize ($10^{28}$\,cm$^2$/s)} & & {\footnotesize(km/s)} & {\footnotesize(km/s/kpc)}& {\footnotesize($\mu$G)}& {\footnotesize(kpc)} & {\footnotesize(kpc)} & &   \\
     \midrule
        I         &  3.0  &  0.5  &  37.3 &   0  & 11.7 & 10 & 2.0 & 1.0,\,1.0 & 0.065 \\ 
        II          &  3.0  &  0.3  &  37.3 &   0  & 11.7 & 10 & 2.0 & 1.0,\,1.0 & 0.084 \\ 
        III        &  5.0  &  0.3  &  37.3 &   0  & 11.7 & 10 & 2.0 & 1.0,\,1.0 & 0.082 \\ 
        IV        &  9.0  &  0.3  &  37.3 &   0  & 11.7 & 10 & 2.0 & 1.0,\,1.0 & 0.062 \\ 
        V         &  9.0  &  0.3  &  37.3 &   0  & 54.7 & 5 & 1.0 & 1.0,\,1.0 & 9$\times 10^{-5}$ \\ 
        VI        &  30.0  &  0.3  &  37.3 &   0  & 54.7 & 5 & 1.0 & 1.5,\,1.5 & 1.4$\times 10^{-5}$ \\ 
        VII       &  30.0  &  0.5  &  37.3 &   0  & 109 & 5 & 1.0 & 1.0,\,1.0 & $<10^{-10}$ \\ 
        VIII      &  30.0  &  0.3  &  37.3 &   1000  & 109 & 5 & 1.0 & 1.0,\,1.0 & $<10^{-10}$ \\ 
        IX        &  30.0  &  0.3  &  37.3 &   0  & 11.7 & 10 & 2.0 & 1.0,\,1.0 & 0.026 \\ 
        X         &  30.0  &  0.5  &  37.3 &   0  & 11.7 & 10 & 2.0 & 1.0,\,1.0 & 0.0082 \\ 
     \midrule
        XI        &  30.0  &  0.5  &  37.3 &   0  & 23.4 & 10 & 1.0 & 1.0,\,1.0 & 0.011 \\ 
        XII        &  100.0  &  0.5  &  37.3 &   0  & 23.4 & 10 & 1.0 & 1.0,\,1.0 & 0.0008 \\ 
        XIII        &  100.0  &  0.5  &  37.3 &   0  & 11.7 & 10 & 1.0 & 1.0,\,1.0 & 0.0002 \\ 
        XIV      &  20.0  &  0.4  &  37.3 &   0  & 11.7 & 10 & 0.5 & 1.0,\,1.0 & 0.017 \\ 
        XV      &  20.0  &  0.45  &  50. &   0  & 11.7 & 10 & 0.5 & 1.0,\,1.0 & 0.015 \\ 
        XVI      &  20.0  &  0.45  &  80. &   0  & 11.7 & 10 & 0.5 & 1.0,\,1.0 & 0.015 \\ 
        XVII      &  20.0  &  0.3/0.5  &  37.3 &   0  & 11.7 & 10 & 0.5 & 1.0,\,1.0 & 0.012 \\ 
        XVIII      &  9.0  &  0.3  &  150. &   0  & 11.7 & 10 & 0.5 & 1.2,\,0.8 & 0.049 \\ 
        XIX      &  15.0  &  0.3  &  200. &   0  & 11.7 & 10 & 0.5 & 1.0,\,1.0 & 0.016 \\ 
        XX (C)    &  9.0  &  0.3  &  150. &   0  & 11.7 & 10 & 0.5 & 1.8,\,0.8 & 0.14 \\ 
      \midrule
        XXI      &  9.0  &  0.3  &  200. &   0  & 11.7 & 10 & 0.5 & 1.5,\,1.5 & 0.083 \\ 
        XXII      &  9.0  &  0.3  &  150. &   0  & 11.7 & 10 & 0.5 & 3.0,\,1.0 & 0.0002 \\ 
        XXIII      &  9.0  &  0.3  &  150. &   0  & 11.7 & 10 & 0.5 & 0.5,\,0.5 & 0.0005 \\ 
        XXIV      &  9.0  &  0.3  &  150. &   0  & 5.8 & 10 & 0.5 & 0.3,\,1.0 & 5$\times 10^{-5}$ \\ 
        XXV      &  9.0  &  0.3  &  20. &   0  & 11.7 & 10 & 0.5 & 1.0,\,1.0 & 0.021 \\ 
        XXVI      &  9.0  &  0.3  &  0. &   0  & 11.7 & 10 & 0.5 & 1.0,\,1.0 & 0.021 \\ 
        XXVII      &  30.0  &  0.3  &  37.3 &   100  & 54.7 & 5 & 1.0 & 1.0,\,1.0 & 0.078 \\ 
        XXVIII      &  30.0  &  0.3  &  37.3 &   500  & 54.7 & 5 & 1.0 & 2.0,\,1.0 & 0.025 \\ 
        XXIX      &  30.0  &  0.3  &  150. &   500  & 54.7 & 5 & 1.0 & 1.0,\,1.0 & 0.043 \\ 
        XXX     &  30.0  &  0.3  &  37.3 &   300  & 54.7 & 5 & 1.0 & 1.8,\,0.8 & 0.057 \\ 
     \midrule
        XXXI      &  9.0  &  0.3  &  37.3 &   0  & 209 & 2.5 & 1.0 & 1.5,\,1.5 & $<10^{-10}$ \\ 
        XXXII      &  9.0  &  0.4  &  37.3 &   0  & 11.7 & 10 & 1.0 & 1.0,\,1.0 & 0.030 \\ 
        XXXIII      &  9.0  &  0.3  &  37.3 &   0  & 74.9 & 2.5 & 1.0 & 1.0,\,1.0 & 8$\times10^{-8}$ \\ 
        XXXIV      &  9.0  &  0.3  &  37.3 &   0  & 74.9 & 2.5 & 1.0 & 1.5,\,1.5 & 5$\times10^{-10}$ \\ 
        XXXV      &  9.0  &  0.3  &  37.3 &   0  & 209 & 2.5 & 1.0 & 1.0,\,1.0 & $<10^{-10}$ \\ 
        \bottomrule
    \end{tabular}
    \caption{The properties of the propagation models considered for the scenarios with two cosmic ray outbursts. The two numbers in the ``ISRF'' column denote the normalizations of the optical and infrared components of the interstellar radiation field, in units relative to the model given in Ref.~\cite{Porter:2005qx}. Finally, the ``$p^{\rm best}$'' column lists the highest $p$-value acquired after fitting the injected spectral indices, spectral cutoffs, and normalizations of the two outbursts (0.1 Myr and 1 Myr old) to the data. Note that the Model XX in this table is identical to Model C discussed throughout the main body of the text.}
    \label{tab:35Models}
\end{table}

\bibliography{bursts.bib}
\bibliographystyle{JHEP}

\end{document}